\documentclass[trackchanges]{aastex701}
\usepackage[normalem]{ulem}

\begin{document}

\title{Detections of Compact Radio Continuum toward Methanol Maser Rings Using the VLA}

\author[orcid=0000-0002-6466-117X,gname=Anna, sname=Bartkiewicz]{Anna Bartkiewicz}
\affiliation{Institute of Astronomy, Faculty of Physics, Astronomy and Informatics, Nicolaus Copernicus University, Gagarina 11, 87-100 Torun, Poland}
\email[show]{annan@astro.umk.pl}  

\author[orcid=0000-0003-4116-4426,gname=olga, sname=Bayandina]{Olga Bayandina} 
\affiliation{SKA Observatory, 2 Fir Street, Black River Park, Observatory, 7925, Cape Town, South Africa}
\email[show]{Olga.Bayandina@skao.int}

\author[orcid=0000-0001-7960-4912, gname=Alberto, sname=Sanna]{Alberto Sanna}
\affiliation{INAF – Osservatorio Astronomico di Cagliari, Via della Scienza 5, 09047 Selargius (CA), Italy}
\email[show]{alberto.sanna@inaf.it}

\author[orcid=0000-0002-1482-8189,gname=Marian, sname=Szymczak]{Marian Szymczak}
\affiliation{Institute of Astronomy, Faculty of Physics, Astronomy and Informatics, Nicolaus Copernicus University, Gagarina 11, 87-100 Torun, Poland}
\email{msz@astro.umk.pl}

\author[orcid=0000-0002-8517-8881, gname=Luca, sname=Moscadelli]{Luca Moscadelli}
\affiliation{INAF – Osservatorio Astrofisico di Arcetri, Largo E. Fermi 5, 50125 Firenze, Italy}
\email{luca.moscadelli@inaf.it}

\author[orcid=0000-0002-1206-9887,gname=Agnieszka, sname=Kobak]{Agnieszka Kobak}
\affiliation{Institute of Astronomy, Faculty of Physics, Astronomy and Informatics, Nicolaus Copernicus University, Gagarina 11, 87-100 Torun, Poland}
\email{akobak@astro.umk.pl}

\author[orcid=0000-0002-0230-5946, gname=Huib, sname= van Langevelde]{Huib Jan van Langevelde}
\affiliation{Joint Institute for VLBI ERIC, Oude Hoogeveensedijk 4, 7991 PD Dwingeloo, The Netherlands}
\affiliation{Sterrewacht Leiden, Leiden University, Postbus 9513, 2300 RA Leiden, The Netherlands}
\email{langevelde@jive.eu}

\author[orcid=0009-0000-0076-1123, gname=Ashwin, sname=Varma]{Ashwin Varma}
\affiliation{Institute of Astronomy, Faculty of Physics, Astronomy and Informatics, Nicolaus Copernicus University, Gagarina 11, 87-100 Torun, Poland}
\email{varma@doktorant.umk.pl}


\begin{abstract}

High-mass protostars are deeply embedded in dust inside their natal cores and are not easily detectable. 
However, maser emission at centimeter wavelengths, owing to its high brightness, enables us to study gas kinematics in protostars' circumstellar regions.
We aim to understand the origin of the ring-like structures outlined by the 6.7~GHz methanol maser emission in six high-mass young stellar objects by performing a sensitive search of the associated radio-continuum emission and derive its properties.  
We used the Karl G.~Jansky Very Large Array in the A configuration at C and K bands in order to image radio-continuum as well as 6.7~GHz methanol and 22~GHz water maser emission. 
We present the first images of the thermal jets towards four targets in our sample, G23.389$+$00.185, G23.657$-$00.127, G28.817$+$00.365, and G30.400$-$00.296. In a further target, G23.207$-$00.377, the complex K band continuum emission makes it unclear whether the detected peaks trace jet knots from a single young protostar or mark multiple compact young protostars. The remaining source G31.047$+$00.356 shows radio continuum emission associated with an evolved H~{\small II} region.
\end{abstract}



\section{Introduction}  \label{sec:intro}

High-mass protostars have large masses (above 8~M$_\odot$) and strongly impact on the surrounding star-forming regions and even on their host galaxies. However, despite their importance, our understanding of how high-mass stars accumulate their masses remains limited. Several factors contribute to this mystery: massive protostars are rare, typically located at great distances, and deeply embedded in the dust of their natal clouds during the early stages of formation.

Fortunately, high-mass young stellar objects (HMYSOs) are marked by a distinctive and invaluable beacon: methanol maser emission at 6.7~GHz \citep[e.g.,][]{menten1991,breen2013}. These masers are known to trace warm gas that is surrounding newly forming high-mass stars, particularly in regions associated with accretion discs, though they are also found in larger envelopes and outflow cavities \citep[e.g.,][]{sanna2010a,devilliers2015}. The presence of 6.7~GHz methanol masers does not only reveal the location of high-mass protostars but it also provides some insights into their evolutionary status \citep{chibueze2017}. For these masers to form, the protostar must be sufficiently luminous to heat the surrounding material to the precise conditions of temperature and density required to excite the methanol molecules, without completely destroying them. In some cases, methanol masers have offered critical clues about the kinematic structure of the system; for instance based on the spatial distribution of the masers, the presence and the orientation in the plane of the sky of accretion discs were inferred \citep[e.g.,][]{sanna2015,bayandina2022}. However, the 6.7~GHz masers typically do not trace the entirety of an accretion disc. Instead, they appear in “patches” that reflect specific physical conditions, often making the interpretation of what they trace challenging. This raises intriguing questions: How does the orientation of high-mass protostar systems affect observational results? How many sources remain undetected simply because their orientation is unfavourable, causing the 6.7~GHz masers to be misaligned with our line of sight? Furthermore, what can we learn from sources with diverse methanol maser distributions, and thus, different accretion disc orientations?

To address the above questions, we searched for methanol masers with ordered structures, particularly those forming ring-like distributions. In such cases, methanol masers may trace entire accretion discs, providing a rare opportunity to clearly study disk kinematics and identify systems with favourable conditions for maser emission.

In 2004$-$2010, we performed a survey of 6.7~GHz methanol maser emission towards 63 HMYSOs using the European VLBI Network (EVN) \citep{bartkiewicz2016}. We analysed the morphology and kinematics of the methanol masers and discovered a new class of sources showing a \textit{ring-like} distribution of maser spots. The most distinctive example of this new type of sources, showing a circularly symmetric distribution of maser emission, was G23.657$-$00.127\footnote{Names are given in Galactic coordinates.} \citep{bartkiewicz2005}. In the sample of 63 sources, we found that a ring-like morphology could be fitted in 17\% of them, an arc-like morphology was found in 8\% of the sample, while 21\% showed maser spots distributed linearly, and in one source only, we detected a single spot. In the remaining 45\% of the sample the maser distribution could not be fitted with simple models. Furthermore, extensive studies of selected targets from the sample of the 
HMYSOs were conducted subsequently. Common processes in massive star formation, such as accretion shocks, thermal and non-thermal emission from proto-stellar jets, as well as free-free emission produced in photo-ionized discs and stellar winds, can result in detectable radio continuum emission at the position of a young massive star \citep{hofner2011}. Therefore, in 2007 we used the Karl G. Jansky Very Large Array (VLA) to search for 8.4~GHz radio continuum counterparts of a subsample of 31 methanol masers that were imaged with the EVN by \citet{bartkiewicz2009}, and in 2009 we searched for 22~GHz water maser emission towards the same subsample 
\citep{bartkiewicz2011}. In 4 out of 31 sources we found relatively weak and compact radio continuum emission that could be physically associated with the methanol emission. Bright water maser emission appeared to be more commonly associated with our targets than radio-continuum emission at a level of several 100~$\mu$Jy~beam$^{-1}$ only, and we obtained 27 positive detections out of 31 targets, with 15 water masers as new discoveries. In 21 out of the 27 targets, both sites of water and methanol masers seemed to be related to the same exciting source. Notably, towards the majority of the ring-like sources, we neither detected radio continuum emission at a 3$\sigma_{\rm rms}$ level of ca.~150~$\mu$Jy~beam$^{-1}$, nor water masers. Therefore, we postulated that methanol maser rings are related to the earliest stages of evolution of the young stars they are associated with, before photo-ionization takes place and before outflows start to hit the circumstellar environment. 

In order to verify this hypothesis, we proposed for higher sensitivity VLA observations to search for ultra- or even hyper-compact radio continuum emission and/or thermal jets towards six targets with ring-like morphology. Taking the HMYSO G23.01$-$0.41 as a reference, methanol maser spots were found there to be excited in the circumstellar envelope around the star, possibly associated with the inner outflow cavity or the surface of a flared disc with the outflow axis lying in the plane of the sky \citep{sanna2015}. By assuming a similar origin for the ring-like sources, we can predict specific geometries that can be tested in our targets: (1) the outflow axis should be found close to the line-of-sight, (2) radio thermal jet emission centred on the methanol maser rings, and (3) the major axis of the jet {\bf is} aligned with the minor axis of a 6.7~GHz maser ellipse (assuming the maser rings are inclined and appear as ellipses).

To follow-up on our previous VLA X-band observations, we selected the C and K bands where the continuum emission should be also dominated by radio free-free, and not contaminated by dust emission. At these frequencies, radio continuum emission from thermal jets and stellar winds at a level of a few hundreds $\mu$Jy/beam (from C to K band) may be expected, following the results of VLA surveys towards massive young stars \citep[e.g.,][]{rosero2016,moscadelli2016,sanna2018,purser2021}. We also took advantage of the improved software correlator (WIDAR) of the Karl G.~Jansky Very large Array (VLA) and observed both 6.7~GHz methanol and 22.2~GHz water masers at the C and K band, respectively, to search for possible weak maser features resolved in the previous VLBI observations. These spectral observations also allow us to obtain maser data at both transitions close in time.  


\section{Sample} \label{sec:sample}

We selected six objects with ring-like morphology of the 6.7~GHz methanol maser emission for which: (1) we have derived the proper motions and (2) the latter were not consistent with rotation motions expected from circumstellar discs \citep{bartkiewicz2024}. A brief summary on each source of the sample is provided below.

{\it G23.207$-$00.377}. The target lies at the distance of 4.18$^{+0.7}_{-0.5}$~kpc as estimated by the trigonometric parallax of 0.239$\pm$0.034~mas \citep{reid2019}. The best fitted ellipse using the procedure by \citet{fitzgibbon99} showed that the 6.7~GHz methanol maser spots follow an ellipse with semi-axes of 129~mas and 40~mas with the position angle (PA) of the major axis of $-$48\degr ~(north to east) \citep{bartkiewicz2024}. The displacements of masers over 10~years showed a clear outward radial motions in the sky with a mean value of 2.61$\pm$0.09~km~s$^{-1}$ \citep{bartkiewicz2024}. The 6.7~GHz maser transition showed little variations in flux density over a decade \citep{szymczak2018} when using the Torun 32~m dish, but the interferometric observations indicated the contrary results - flux densities varied for majority of the maser cloudlets \citep{bartkiewicz2024}. The 22~GHz water maser emission was also found towards this methanol ring and we stated the disc-outflow scenario relating to the methanol and water maser distributions, respectively \citep{bartkiewicz2011}.

{\it G23.389$+$00.185}. The derived trigonometric parallax is 0.208$\pm$0.025~mas corresponding to a distance of 4.8$^{+0.7}_{-0.5}$~kpc \citep{reid2019}. High-resolution near- and mid-infrared imaging by \citet{debuizer2012} revealed nearby bright source with a peak located northward (ca.~0.25'') from the methanol maser ring. \citet{hu2016} reported weak continuum emission at 6~GHz with the integral flux S$_{\rm int}$=0.38~mJy that showed an elongated structure and may be associated with a jet. The 6.7~GHz methanol maser spots follow an ellipse with 105 and 51~mas semi-axes, PA=$+$52\degr ~\citep{bartkiewicz2024}. The proper motion studies revealed that, again, the outward radial motions dominate with a mean value of 3.4$\pm$0.4~km~s$^{-1}$ \citep{bartkiewicz2024}. Also, the single dish monitoring of the 6.7~GHz methanol maser line showed little variation of the flux density \citep{szymczak2018}, while the interferometric data suggest changes in the spectra \citep{bartkiewicz2024}. Moreover, the target has been reported as undergoing an outburst recently \citep{tanabe2023}. Two 22~GHz water maser groups were detected in association with the methanol ring \citep{bartkiewicz2011}.

{\it G23.657$-$00.127}. The methanol masers in this source are distributed almost circularly, the best-fitted ellipse has an eccentricity of 0.38. The semi-axes are 136 and 124~mas with PA=$-$19\degr ~\citep{bartkiewicz2020}, and the distance to the target is 3.19$^{+0.46}_{-0.35}$~kpc \citep{bartkiewicz2008}. The methanol masers have expanded mainly in the radial direction outward from the centre with a mean velocity of 3.2~km~s$^{-1}$. Considering the proper motions of the masers and near- and mid-infrared images \citep{debuizer2012}, we suggested the association of the methanol ring with a spherical outflow arising from an (almost) edge-on disc, or a wide angle wind at the base of a protostellar jet (see Fig. 7 in \citet{bartkiewicz2020}). No water emission was detected above a 5$\sigma$ level of 15-25~mJy \citep{bartkiewicz2011}.

{\it G28.817$+$00.365}. The methanol maser spot distribution can be fitted by an ellipse with semi-axes of 60 and 27~mas with PA of $+$53\degr. Towards this target we detected radio continuum at 8.4~GHz using the VLA in 2007 \citep{bartkiewicz2009}. The detected continuum emission was relatively weak with the peak flux (S$_{\rm p}$) of 0.81~mJy~beam$^{-1}$ and S$_{\rm int}$=0.8~mJy, and compact with a size of ca.~0.6''$\times$0.5''. The methanol maser emission lies close (ca.~80~mas) to the peak continuum emission that likely points to the location of the central object. The kinematic distance to this target is estimated as 4.6~kpc \citep {reid2019}. The proper motions of the 6.7~GHz methanol maser spots over 8~yr range from 1.6 to 13~km~s$^{-1}$ and suggest a combination of rotation and expansion motions \citep{bartkiewicz2024}. Again, the interferometric observations revealed variability of the flux densities of masing regions \citep{bartkiewicz2024}, contrary to the single-dish data \citep{szymczak2018}. The 22~GHz water maser emission was found on the west side of the radio-continuum \citep{bartkiewicz2011}. 

{\it G30.400$-$00.296}. In this source only four groups of methanol masers were detected, which limits the reliability of fitting an ellipse to the maser positions \citep{bartkiewicz2009}. The proper motions over 8~yr are directed towards the west direction with typical velocities of 3.5~km~s$^{-1}$ assuming the far kinematic distance of 6.4~kpc \citep{reid2019}. The 22~GHz water masers were found 10\farcs5 away from the methanol ones \citep{bartkiewicz2011}. 

{\it G31.047$+$00.356}. The methanol maser emission also consists of four maser groups with a  significant extension of the western emission; the best-fitted ellipse has major and minor axes of 34 and 14~mas with PA of $+$63\degr. The proper motion measurements over 8~yr range from 0.7 to 7.1~km~s$^{-1}$ for a far kinematic distance of 4.9~kpc \citep{reid2019} and are directed in both radial and tangential, directions \citep{bartkiewicz2024}. \citet{hu2016} reported tentative 6~mJy continuum emission that likely coincides with the methanol emission; that emission was not detected earlier by \citet{bartkiewicz2009}. Water masers were found 4.6" away from the methanol masers \citep{bartkiewicz2011}.

\section{Observations and data reduction\label{sec:obs}}
\subsection{Very Large Array}

\begin{deluxetable*}{ccccccccc}
\tablenum{1}
\tablecaption{Details of observations\label{tab:obs}}
\tablewidth{0pt}
\tablehead{
\colhead{Target} & \multicolumn2c{Continuum 6~GHz}&  \multicolumn2c{Meth. masers}& \multicolumn2c{Continuum 22~GHz}&  \multicolumn2c{Water masers}\\
\colhead{Gll.lll$+$bb.bbb} & \colhead{Synth. beam} & \colhead{1$\sigma$} & \colhead{Synth. beam} &\colhead{1$\sigma$} & \colhead{Synth. beam} & \colhead{1$\sigma$} & \colhead{Synth. beam} &\colhead{1$\sigma$} \\
&\colhead{("$\times$"; \degr)}& \colhead{($\mu$Jy~b$^{-1}$)} & \colhead{("$\times$"; \degr)}& \colhead{(mJy~b$^{-1}$)} & \colhead{("$\times$"; \degr)}& \colhead{($\mu$Jy~b$^{-1}$)} & \colhead{("$\times$"; \degr)}& \colhead{(mJy~b$^{-1}$)} 
}
\decimalcolnumbers
\startdata
G23.207$-$00.377 & 0.623$\times$0.350;$-$37 & 8 & 0.434$\times$0.269; $-$30 & 11 & 0.172$\times$0.102; $+$24 & 7 & 0.144$\times$0.087; $+$25 & 3 \\
G23.389$+$00.185 & 0.579$\times$0.356;$-$35 & 7 & 0.413$\times$0.263; $-$26 & 10 & 0.146$\times$0.095; $+$6 & 6 & 0.100$\times$0.072;  $+$0 & 3\\
G23.657$-$00.127 & 0.558$\times$0.356;$-$33 & 7 & 0.399$\times$0.268; $-$26 & 10 & 0.159$\times$0.098; $+$28 & 7 & 0.108$\times$0.080; $+$10 & 5\\
G28.817$+$00.365 & 1.095$\times$0.368;$-$47 & 10 & 0.632$\times$0.273; $-$50 & 14 & 0.189$\times$0.104; $-$48 & 13 & 0.123$\times$0.079; $-$44 & 5\\
G30.400$-$00.296 & 0.959$\times$0.362;$-$46 & 12 & 0.562$\times$0.279; $-$50 & 13 & 0.190$\times$0.099; $-$44 & 5 & 0.122$\times$0.078; $-$42 & 3\\
G31.047$+$00.356 & 0.875$\times$0.371;$-$44 & 8 & 0.507$\times$0.276; $-$46 & 12 & 0.203$\times$0.103; $-$48 & 7 & 0.131$\times$0.085; $-$45 & 5\\
\enddata
\end{deluxetable*}

The VLA observations of the source sample were conducted in eight sessions during the  period of March-May 2018 under the project code 18A-032. The observations were made in A-configuration, i.e. with the highest possible resolution of the array. The first two sessions were devoted to the C-band observations (22 and 26 March 2018), while in the remaining 6 sessions (9, 19, 21, 27, 29 April and 3 May 2018) the K-band receiver was used. The C-band blocks had duration of 1 hour and contained three target sources each ($\sim$6.5~min on-source per scan, giving the total on-source time of $\sim$13~min); the K-band blocks had duration of 1~h~20~min and were prepared for each individual source of the sample ($\sim$2~min on source per scan, giving the total on-source time of $\sim$38~min). The observations were made in phase-reference mode. 3C286 (J1331$+$3030) was used as flux and bandpass calibrator. Continuum emission was observed with 128 1-MHz channels per spectral window. In C band we used 30 spectral windows covering the frequencies from ca.~4~GHz till 8~GHz and in K band 60 spectral windows from ca.~18~GHz till 26~GHz. Maser lines were observed in narrower spectral windows with 4096 channels of 2 kHz in C band and 256 channels of 30 kHz in K band. Observation parameters, including the synthesized beam size and the rms noise level, for continuum and spectral line data, are presented in Table~\ref{tab:obs}. 

The post-correlation data reduction was performed using Common Astronomy Software Applications (CASA ver. 5.6, \cite{McMullin07}). The basic flagging and data calibration was done with the VLA CASA Calibration Pipeline. The calibrated data for each target source was imaged using the Clark CLEAN algorithm \citep{1980AA....89..377C}. The imaging process was automated with Python scripts prepared for our data. The pixel sizes of the images were 0\farcs06 and 0\farcs02 at the C and K bands, respectively. The synthesised beamsizes are listed in Table~\ref{tab:obs}.  

A two-dimensional Gaussian brightness distribution was fit to the emission peak in case of the continuum emission and to every channel of the maser images using the JMFIT task of the Astronomical Image Processing System package, AIPS (NRAO 2024). The fitting cut-off level was set at the flux density above 3$\sigma$. Each fit determines the absolute position (with formal uncertainties), the flux density and brightness of either the continuum emission or single maser spots. 

\subsection{Torun 32-m radiotelescope\label{sec:torunmon}}
A large sample of the 6.7~GHz methanol masers in our Galaxy has been monitored with the 32-m Torun radio telescope in the period 2009-2013 as described in details in \cite{szymczak2018}. Follow-up observations have been ongoing since early 2018 \citep{wolak2024}. The spectral resolution has been  0.09~km~s$^{-1}$ after Hanning smoothing, a typical sensitivity (3$\sigma$) has been estimated as  0.8\,Jy, and the flux density calibration accuracy as 10\%.

\section{Results\label{sec:res}}

\begin{deluxetable*}{cccccccc}
\tablenum{2}
\tablecaption{Detection of continuum emission\label{tab:rescont}}
\tablewidth{0pt}
\tablehead{
\colhead{Target} & \multicolumn2c{Coordinates} & \colhead{Distance} & \multicolumn2c{Continuum C band} &  \multicolumn2c{Continuum K band} \\
\colhead{Gll.lll$+$bb.bbb} & \colhead{RA (J2000)} & \colhead{Dec (J2000)} & & \colhead{S$_p$} & \colhead{S}& \colhead{S$_p$} & \colhead{S} \\ 
& \colhead{(h:m:s)} & \colhead{($^{\rm o}$:':")} & \colhead{(kpc)} & \colhead{(mJy~b$^{-1}$)} & \colhead{(mJy)} & \colhead{(mJy~b$^{-1}$)} & \colhead{(mJy) }
}
\decimalcolnumbers
\startdata
G23.207$-$00.377 & 18:34:55.203 & $-$08:49:14.888  & 4.2$^{+0.7}_{-0.5}$ & 0.106 & 0.137 &  \\
             VLA-1& 18:34:55.202 & $-$08:49:14.879  &                     & & & 0.091 & 0.070 \\
             VLA-2& 18:34:55.208 & $-$08:49:14.961  &                     & & & 0.097 & 0.086 \\
             VLA-3& 18:34:55.193 & $-$08:49:14.982  &                     & & & 0.091 & 0.088 \\
             VLA-4& 18:34:55.203 & $-$08:49:15.219  &                     & & & 0.042 & 0.037 \\
G23.389$+$00.185 & 18:33:14.319 & $-$08:23:57.548  & 4.8$^{+0.7}_{-0.5}$ & 0.125 & 0.099 \\
                 & 18:33:14.319 & $-$08:23:57.614  &                     & & & 0.139 & 0.200  \\
G23.657$-$00.127 & 18:34:51.565 & $-$08:18:21.436  & 3.2$^{+0.5}_{-0.4}$ & 0.097 & 0.068 & 0.021 &   \\
G28.817$+$00.365 & 18:42:37.346 & $-$03:29:41.002  & 4.6$\pm$0.5$^k$ & 0.437 & 0.470 & \\
                 & 18:42:37.343 & $-$03:29:40.919 &                  & & & 1.057 & 1.153 \\
G30.400$-$00.296 & 18:47:52.299 & $-$02:23:15.980 & 6.4$\pm$1.5$^k$ & 0.036 & & 0.066 & 0.096 \\
G31.047$+$00.356 & 18:46:43.887 & $-$01:30:53.279 & 5.0$\pm$0.8$^k$ & 0.106 & 4.310 & 0.018   \\
\enddata
\tablecomments{Distances (Col.~4) determined via trigonometric measurements \citep[][]{reid2019,bartkiewicz2008} or kinematic distances (marked by the $k$ index) using the calculator from \citet{reid2019}.}
\end{deluxetable*}

\begin{deluxetable*}{ccccccccc}
\tablenum{3}
\tablecaption{Detection of methanol and water maser emission\label{tab:resmaser}}
\tablewidth{0pt}
\tablehead{
\colhead{Target} & \multicolumn2c{Coordinates} & \multicolumn5c{Masers} &  Notes\\
\colhead{Gll.lll$+$bb.bbb} & \colhead{RA (J2000)} & \colhead{Dec (J2000)} & \colhead{S$_p$} & \colhead{V$_p$}& \colhead{$\Delta$V} & \colhead{V$_{min}$} & \colhead{V$_{max}$}& \\
& \colhead{(h:m:s)} & \colhead{($^{\rm o}$:':")}  & \colhead{(Jy~b$^{-1}$)} & \colhead{(km~s$^{-1}$)} & \colhead{(km~s$^{-1}$)} &\colhead{(km~s$^{-1}$)} & \colhead{(km~s$^{-1}$)} &  
}
\decimalcolnumbers
\startdata
G23.207$-$00.377 & 18:34:55.21094 & $-$08:49:14.9183 & 26.128 & 77.25 & 13.170 & 72.25 & 85.42 & Methanol \\
                 & 18:34:55.20902 & $-$08:49:14.9455 & 9.412 & 79.37 & 7.16 & 72.21 & 79.37 & Water \\
G23.389$+$00.185 & 18:33:14.32228 & $-$08:23:57.5146 & 32.028 & 75.29 & 5.97 & 71.60 & 77.57 & Methanol  \\
                 & 18:33:14.31685 & $-$08:23:57.6620 & 36.019 & 74.89 & 2.10 & 73.63 & 75.73 & Water\\
G23.657$-$00.127 & 18:34:51.56170 & $-$08:18:21.3413 & 12.557 & 82.44 & 14.57 & 73.66 & 88.23 & Methanol \\
                 & & & $<$0.015 & & & & & water\\
G28.817$+$00.365 & 18:42:37.35000 & $-$03:29:41.0002 &  5.359 & 91.11 & 5.89 & 87.51 & 93.39 & Methanol\\
                 & 18:42:37.30070 & $-$03:29:41.3500 &  61.206 & 86.17  & 47.61 & 47.41 & 95.02 & Water \\
G30.400$-$00.296 & 18:47:52.30373 & $-$02:23:15.9641 &  3.600 & 105.01& 8.07 & 97.73 & 105.80 & Methanol \\
                 & 18:47:52.30339 & $-$02:23:15.9868 &  0.494 & 107.74 & 17.27 & 90.89 & 108.16 & Water\\
G31.047$+$00.356 & 18:46:43.84965 & $-$01:30:54.2185 &  5.029 & 82.55 & 9.13 & 77.81 & 86.94 & Methanol\\
                 & 18:46:43.53980 & $-$01:30:52.8632 &  0.344 & 77.86 & 8.43 & 77.01 & 85.44 & Water\\
\enddata
\end{deluxetable*}

Detections of continua at C and K bands are summarised in Table \ref{tab:rescont}. We list the coordinates (RA, Dec) of each continuum peak of emission, its peak flux density (S$_p$) and flux density (S). In a case of non-detection, we list the upper limit of the emission (the 3$\sigma$ level). Imaging of the 6.7~GHz methanol and 22~GHz water maser transitions is summarised in Table \ref{tab:resmaser}. Maser spots are described by the following parameters: the coordinates (RA, Dec), intensities (S$_p$), the LSR velocities (V$_p$) of the brightest spots at each transition, and the velocity ranges of emission ($\Delta$V, V$_{min}$ and V$_{max}$). The overall results of the new data are presented in Figures \ref{fig:G23p207}--\ref{fig:G31p047}.

We detected radio continuum in all of our targets using the high-sensitivity of the VLA. The 6~GHz continuum emission is clearly detected in five sources (83\% of the sample) and the 22~GHz emission in four sources (67\% of the sample). The continuum sources appeared relatively weak with a range of peak flux densities from 0.097~mJy~b$^{-1}$ to 0.437~mJy~b$^{-1}$ and from 0.042~mJy~b$^{-1}$ to 1.057~mJy~b$^{-1}$ at the C and K band, respectively. In three cases, where both radio continua were imaged, the higher frequency band, i.e. K band, shows the brighter emission. 

Moreover, we detected both the 6.7~GHz methanol and 22~GHz water emission in all targets but G23.657$-$00.127; this target showed only the methanol masers. In Figures \ref{fig:G23p207}--\ref{fig:G31p047} we present the summary of detections and also compare with previous observations \citep{bartkiewicz2009, bartkiewicz2011}. 

The monitoring program using the single-dish allows us to present the dynamic spectra of the 6.7~GHz methanol maser variability in our six targets over 15~years (Figure \ref{fig:monitoring} in the Appendix).

\subsection{Individual targets\label{subsec:targets}}
\begin{figure*}
\gridline{\fig{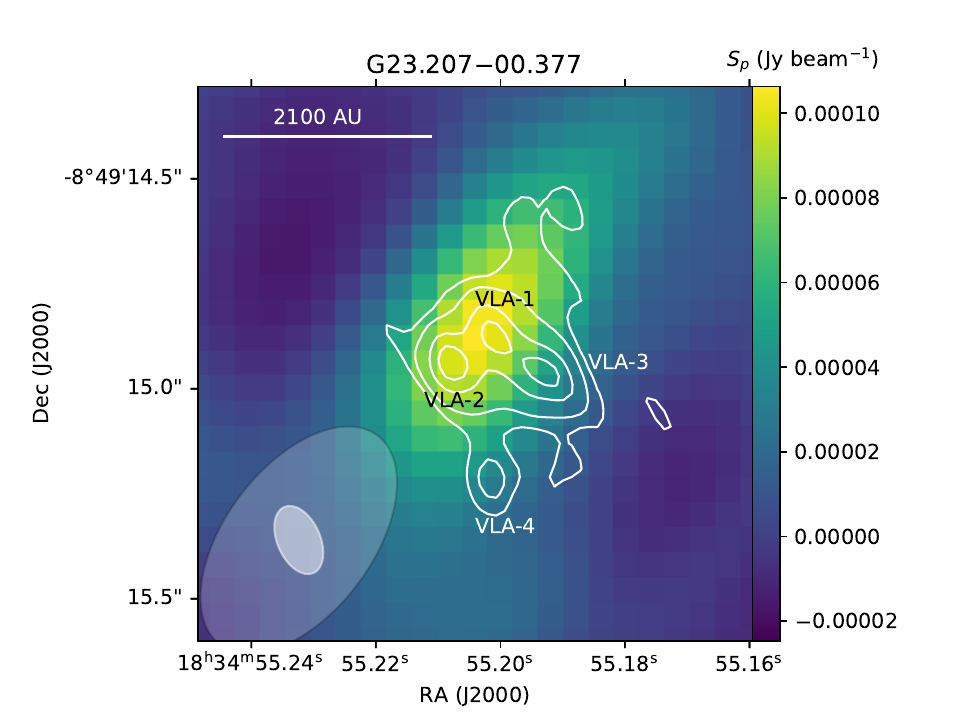}{0.5\textwidth}{(a)}
          \fig{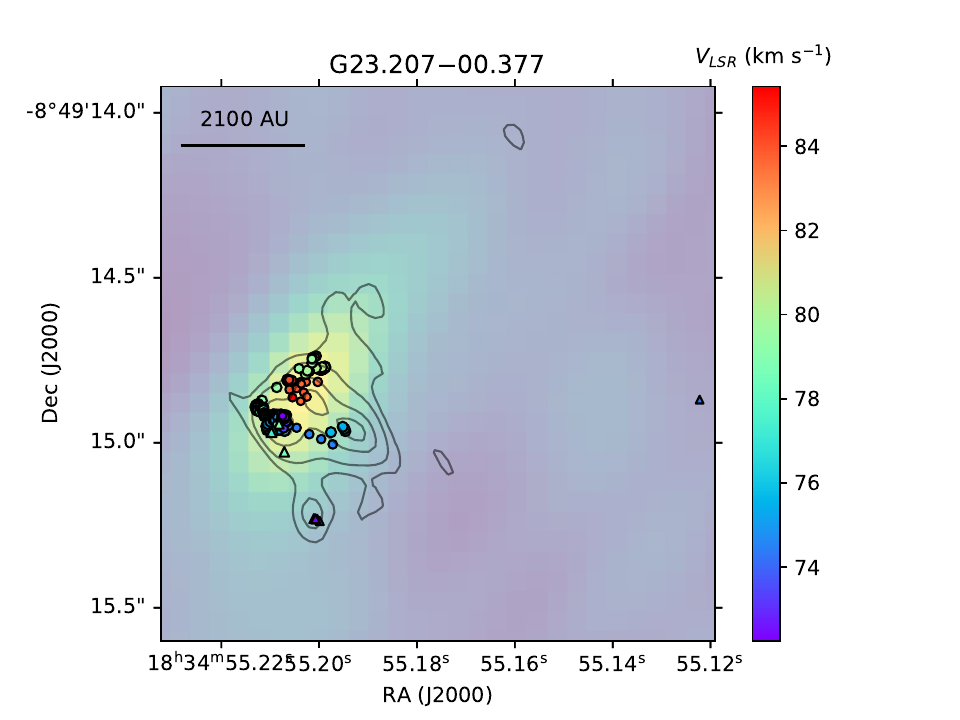}{0.5\textwidth}{(b)}
          }
\gridline{\fig{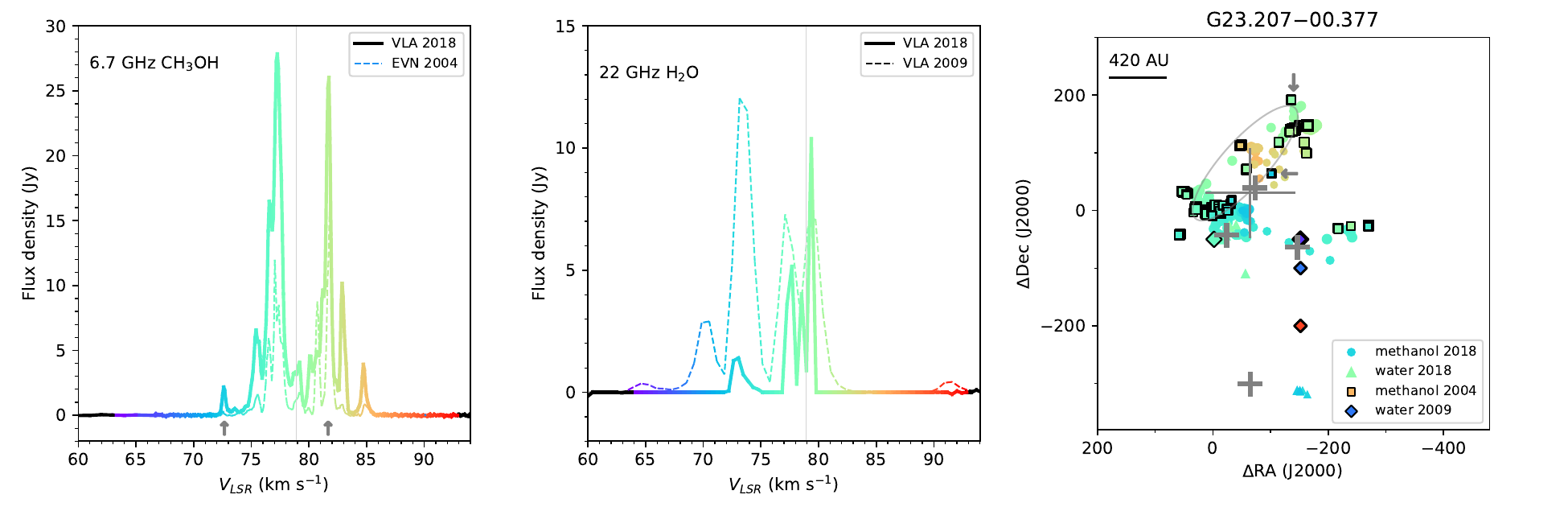}{1\textwidth}{(c)}
          }
\caption{G23.207$-$00.377: (a) The continuum emission detected using JVLA at the C band (colors) and K band (white contours). The contours correspond to 3$\sigma$, 7$\sigma$, 9$\sigma$, and 12$\sigma$ (where $\sigma$ is listed in Table \ref{tab:obs}) The synthesized beam sizes are presented at the bottom, left corner; the transparent ellipse with black outline and the white ellipse correspond to C- and K-band data, respectively. (b) The continuum emission overlaid with methanol and water maser spots detected using JVLA. Their sizes are proportional to the logarithm of S$_p$ of methanol and water peak fluxes, respectively. The colors of spots correspond to the LSR velocities as presented on wedges. (c) The spectra and the distributions of the maser transitions presented in this publication and from \citet{bartkiewicz2009, bartkiewicz2011}. The (0,0) point corresponds to the brightest methanol maser spot (Table~\ref{tab:resmaser}). The vertical light gray lines in spectra represent the systemic velocities of sources as in \citep{szymczak2007}. The sizes of the symbols are proportional to the logarithm of the peak flux density of a spot. The gray arrows at the left and right panels point out the variable (over 14~yr) methanol maser features. The thin and thick crosses corresponds to the peaks of C- and K-band radio continua with their positional uncertainties, respectively. The gray ellipse traces the best flux-weighted fit as in \citep{bartkiewicz2024}.
\label{fig:G23p207}}
\end{figure*}

{\it G23.207$-$00.377}. We detected unresolved continuum emission at 6~GHz and complex 22~GHz emission consisting four peaks (Figure \ref{fig:G23p207}). Three peaks lie within a region of 0\farcs3$\times$0\farcs2 and their SNRs are above 12; two of them, which we name as VLA-1, VLA-2, the brightest ones (S$_p$ of 0.091 and 0.097~mJy~b$^{-1}$), the third one, VLA-3, is located westward. The forth peak, VLA-4, is the weakest source with S$_p$ of 0.042~mJy~b$^{-1}$ and is detected to the south with SNR=6. The C band peak is consistent with the VLA-1 position and it differs by $\sim$100~mas from the VLA-2 position.

The methanol maser spots coincide with the radio continuum, with the C-band peak and three brightest K-band peaks VLA-1, VLA-2 and VLA-3. The water masers are found closely, separated by ca.~0\farcs1-0\farcs3 from the methanol ring and in majority on the SW side at distances of 0\farcs5--{\bf 1\arcsec}. That is similar as it was reported by \citet{bartkiewicz2011}. When comparing the methanol maser spectra one can notice that the VLBI resolved the emission and also two spectral features at the LSR velocities of 72.5~km~s$^{-1}$ and 81.5~km~s$^{-1}$ show variability over 14~years. They are clearly related to the methanol ring as imaged earlier using EVN (Figure \ref{fig:G23p207}). Moreover, we note that the compact methanol maser emission is related to the VLA-1 radio continuum and the SW blue-shifted masers are related to the VLA-3. There is also emission towards the VLA-2, which has to be extended emission that was filtered out using the VLBI. However, since the methanol maser spots detected using VLA are continuously distributed between radio continua VLA-1,2,3, while the EVN imaged only the separated groups of masing regions related to the radio continuum peaks, we can not clearly distinguish, if we have recovered resolved emission or spurious VLA spots result from the emission of closely located methanol masers. Thanks to the Torun monitoring program, we noticed a brightening of spectral features between the LSR velocities from 80 and 82~km~s$^{-1}$ from the end of 2023 (Figure \ref{fig:monitoring} in the Appendix); this region is likely related to the northern part of the fitted ellipse. Concerning the water maser transition, two new spectral features are found at the most blue-shifted LSR velocities ($<$73~km~s$^{-1}$). One coincides with the south VLA-4 continuum peak and the second lies 1" ~westward from the methanol maser ring without any radio counterpart. 

\begin{figure*}
\gridline{\fig{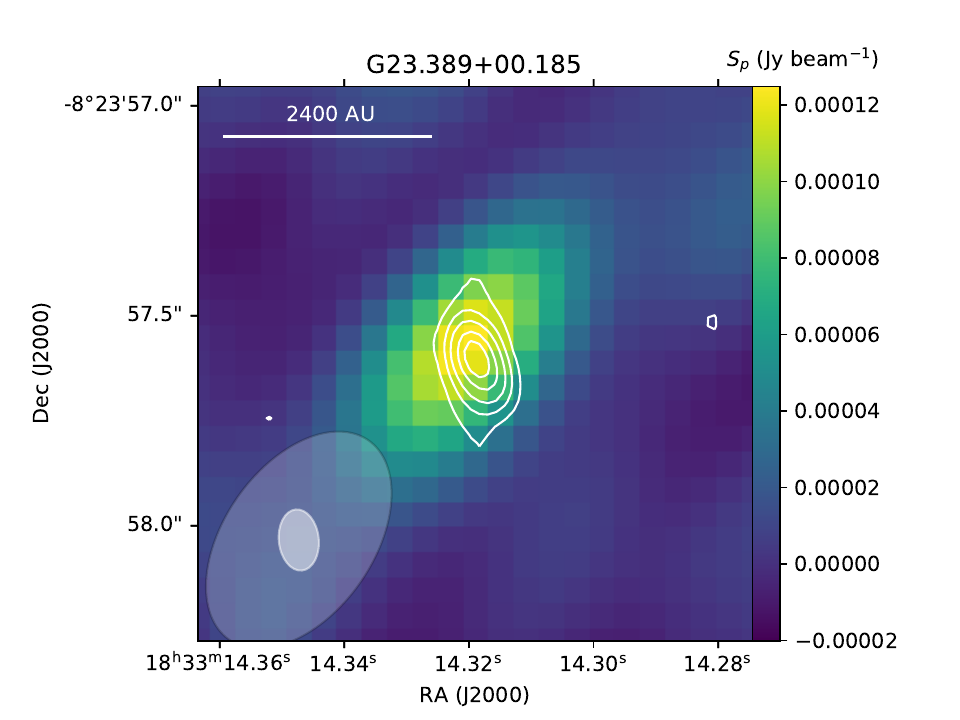}{0.5\textwidth}{(a)}
          \fig{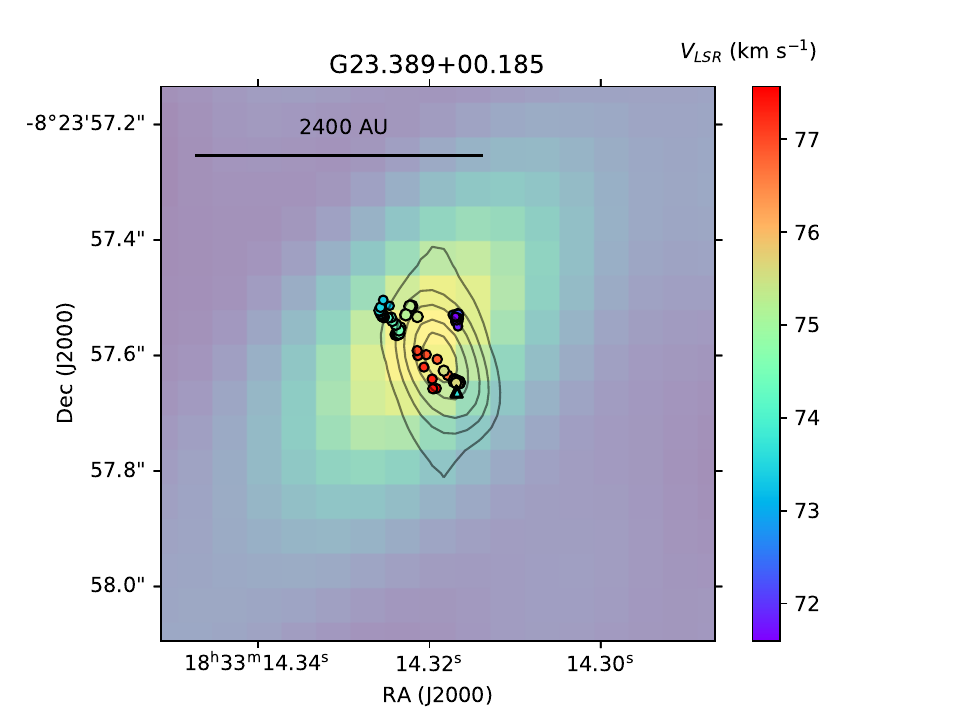}{0.5\textwidth}{(b)}
          }
\gridline{\fig{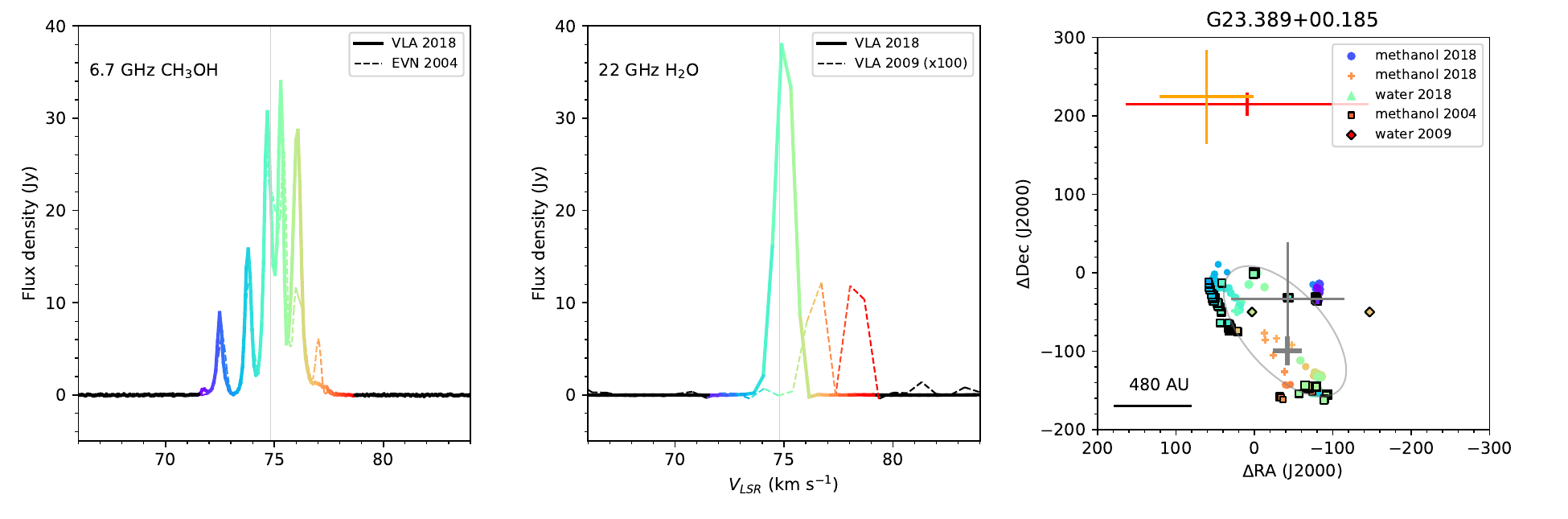}{1\textwidth}{(c)}
          }
\caption{Similar as Figure\ref{fig:G23p207} but for G23.389$+$00.185. (a) The contours of the K-band emission correspond to 3$\sigma$, 7$\sigma$, 11$\sigma$, 16$\sigma$, and 20$\sigma$. (c) The red and orange crosses trace the NIR and MIR emission from \citep{debuizer2012}, respectively. The crosses trace the likely spurious spots (see Sect.~\ref{subsec:targets})
\label{fig:G23p389}}
\end{figure*}

{\it G23.389$+$00.185}. Unresolved continua at the 6~GHz and 22~GHz bands have been detected; their peaks coincide within 0\farcs07 (Figure \ref{fig:G23p389}). The NS elongation at the higher frequency band is tentative, extended imaging with round beamsize do not confirm such characteristic. The masers at both transitions are distributed projected on these continua. The methanol maser spot distribution and the spectra are similar as previously detected using EVN, indicating that there are not any extended emission. However, we note that spurious spots (mentioned also in G23.207$-$00.377 as a results of closely located masers not resolved by the VLA beam) are likely to appear as linear V$_{\rm LSR}$ gradients of weak spots connecting intense masers at different LSR velocities; we point them at Figure \ref{fig:G23p389} based on the intensities. We notice that the spectral feature at the LSR velocity of 77~km~s$^{-1}$ lowered in brightness what is confirmed by the monitoring program. Also, a brightening of spectral features between the LSR velocities from 74.3 and 75.5~km~s$^{-1}$ from the end of 2023 is seen using single-dish data (Figure \ref{fig:monitoring} in the Appendix). They are related to the northern parts of the methanol ring. Considering the water maser transition, we notice significant variability over nine years (Figure \ref{fig:G23p389}), the spectrum and the maser spot distribution are relocated in the LSR velocity domain and on the sky indicating that different regions started maser pumping. 

\begin{figure*}
\gridline{\fig{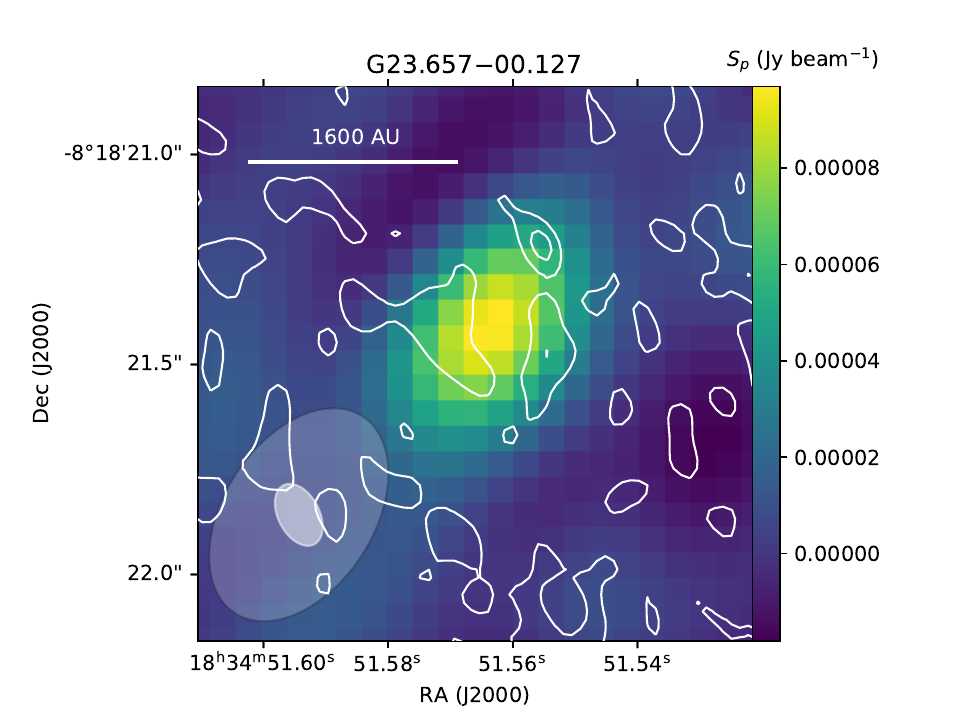}{0.5\textwidth}{(a)}
          \fig{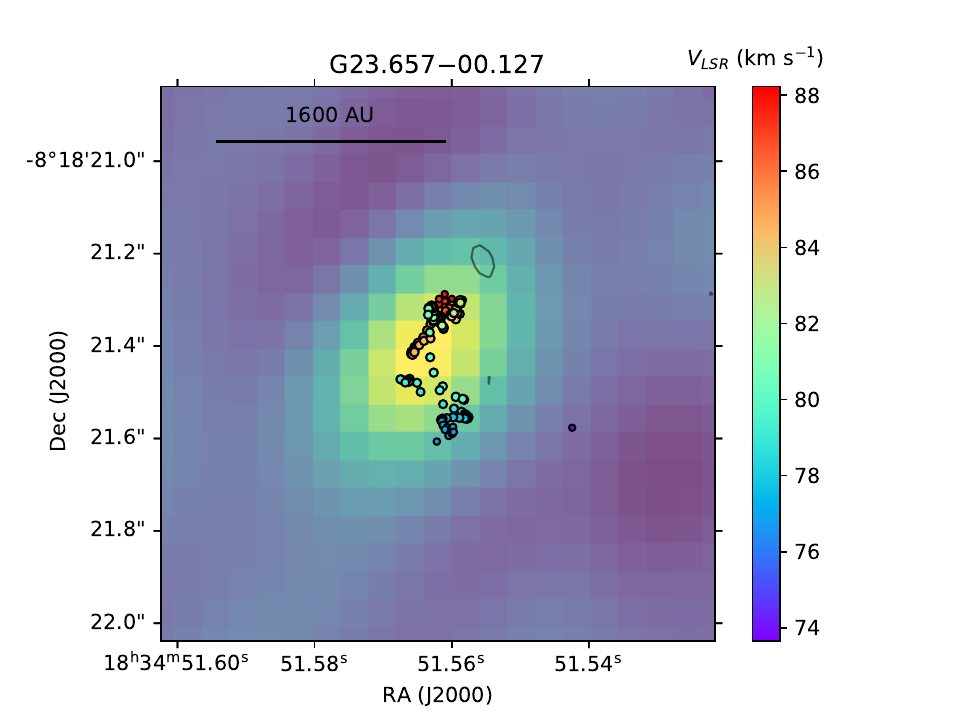}{0.5\textwidth}{(b)}
          }
\gridline{\fig{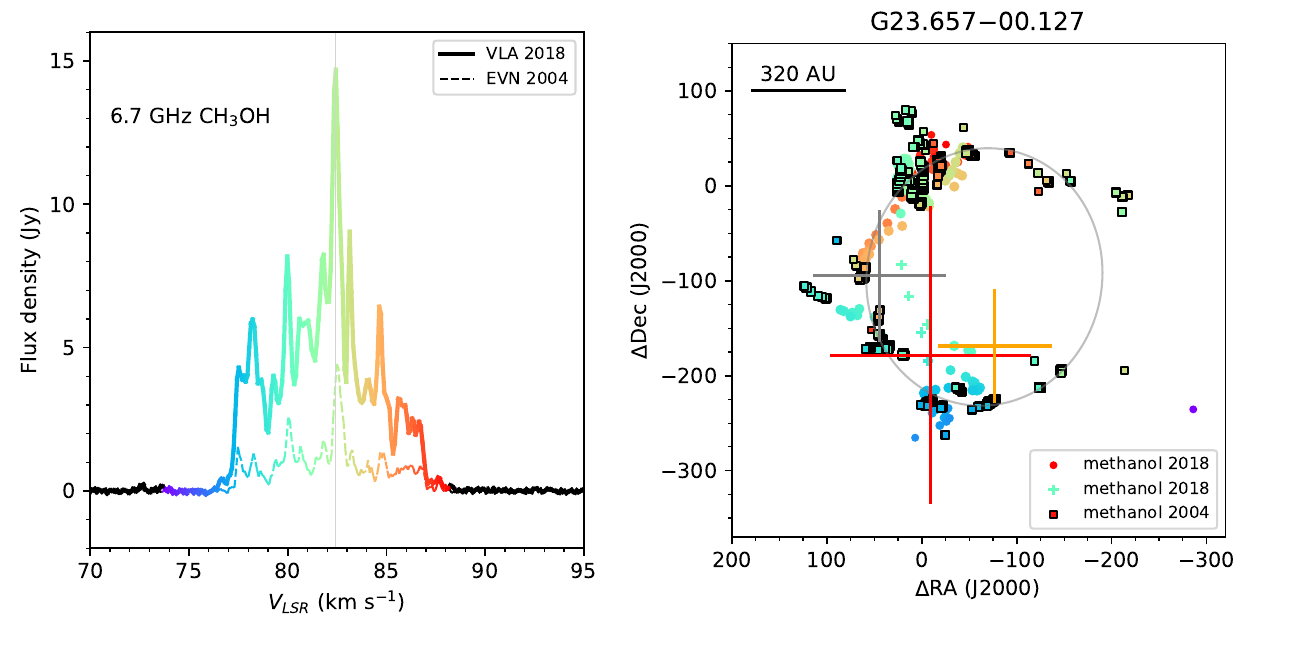}{0.8\textwidth}{(c)}
          }
\caption{Similar as Figure \ref{fig:G23p207} but for G23.657$-$00.127: (a) The contours of the K-band emission correspond to 1$\sigma$ and 3$\sigma$. (c) The red and orange crosses trace the NIR and MIR emission from \citep{debuizer2012}, respectively. The crosses trace the likely spurious spots (see Sect.~\ref{subsec:targets}).
\label{fig:G23p657}}
\end{figure*}

{\it G23.657$-$00.127}. We detected the C-band radio emission with a peak of flux density of 97~$\mu$Jy~beam$^{-1}$ and none radio continuum was detected at 22~GHz with a limit of 21~$\mu$Jy~beam$^{-1}$ (3$\sigma$) (Figure \ref{fig:G23p657}). Similarly, we detected methanol maser emission and no water masers with a limit of 15~mJy~beam$^{-1}$ (3$\sigma$). This is in agreement with previous studies \citep{bartkiewicz2011}. The methanol maser spots recovered using VLA roughly follow the ring-like structure imaged using EVN \citep{bartkiewicz2005} and the whole masing region coincides with the peak of the radio continuum. This target shows the lowest variability index among our sample (Figure \ref{fig:monitoring} in the Appendix). We note, that the inner spots (marked in Figure \ref{fig:G23p657}) may be the result of closely located masers not resolved by the VLA beam as in G23.389$+$00.185. 

\begin{figure*}
\gridline{\fig{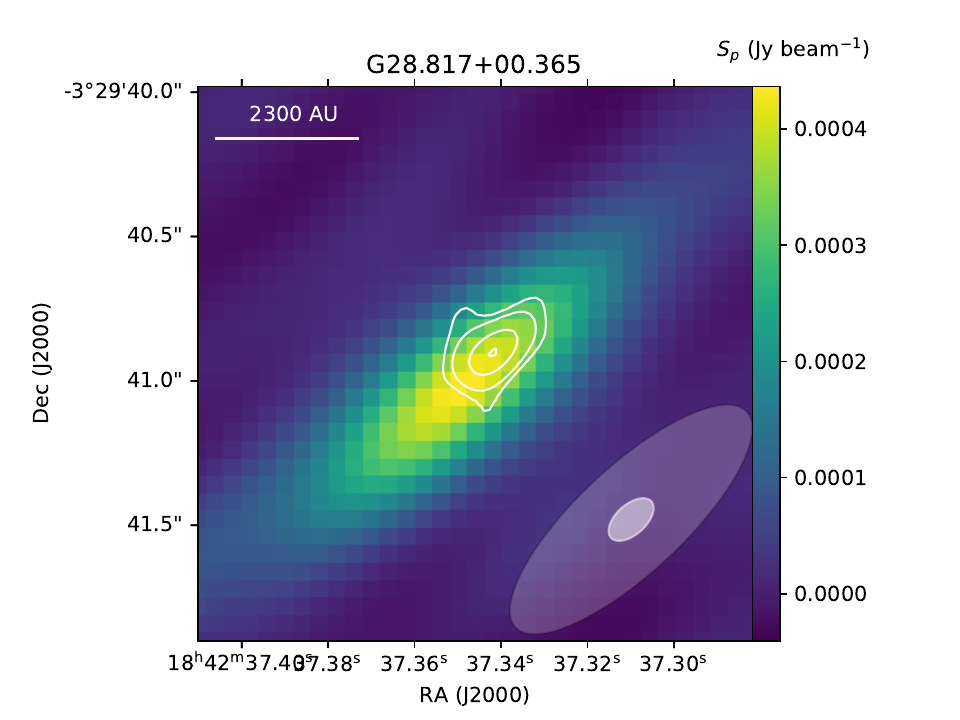}{0.5\textwidth}{(a)}
          \fig{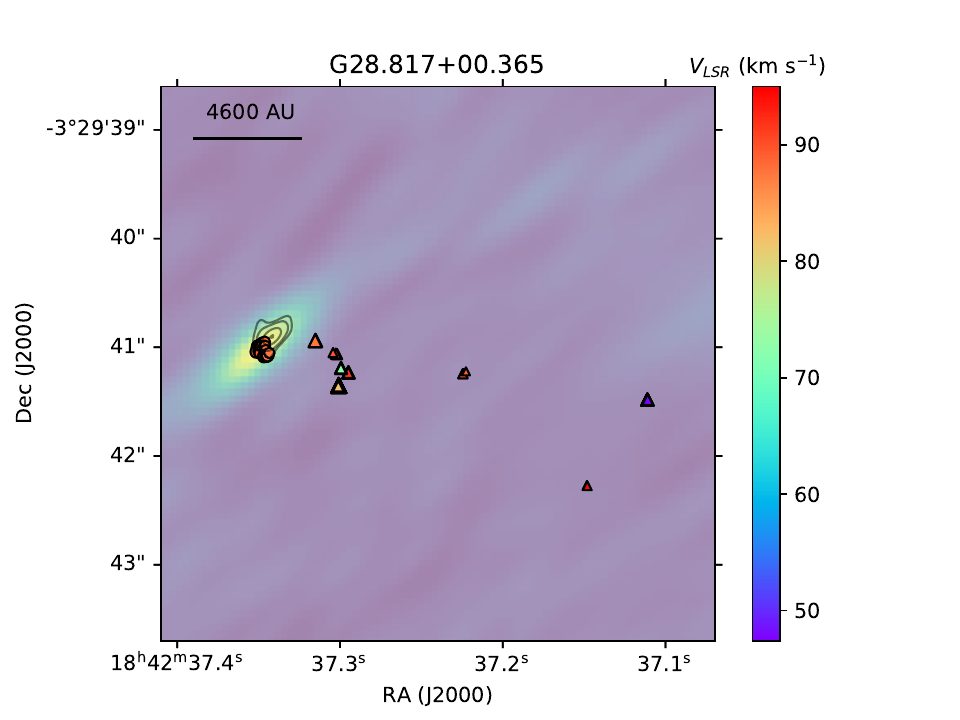}{0.5\textwidth}{(b)}
          }
\gridline{\fig{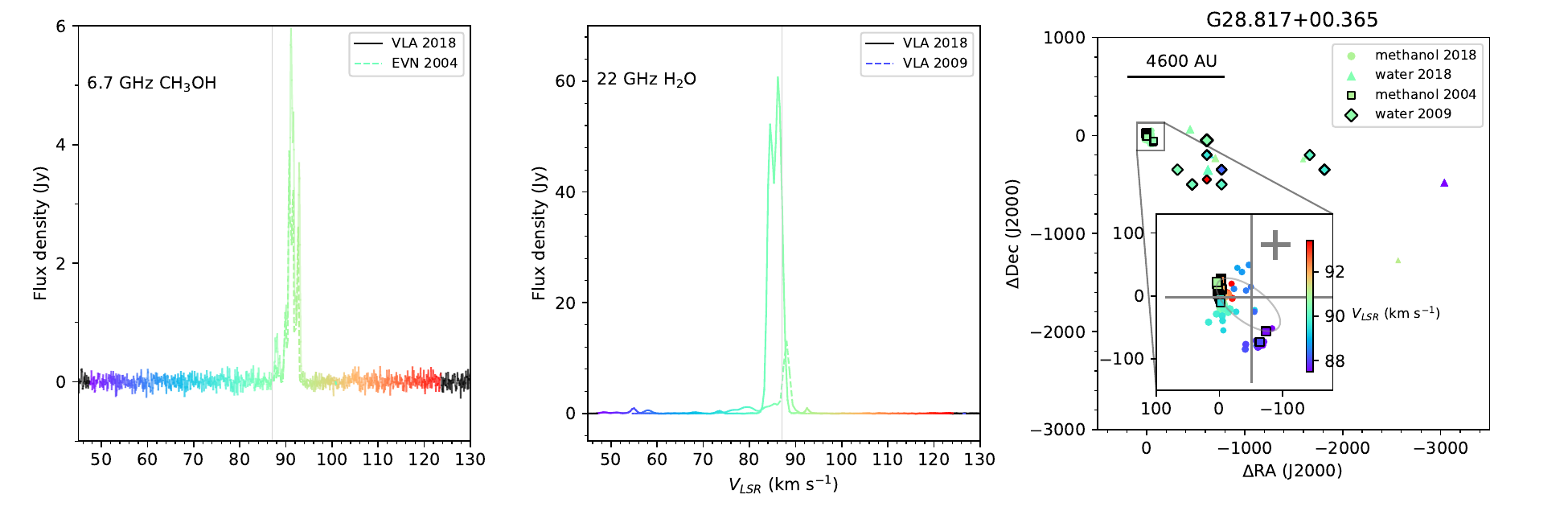}{1\textwidth}{(c)}
          }
\caption{Similar as Figure \ref{fig:G23p207} but for G28.817$+$00.365: (a) The contours of the K-band emission correspond to 3$\sigma$, 10$\sigma$, 40$\sigma$, and 80$\sigma$. 
\label{fig:G28p817}}
\end{figure*}

\begin{figure*}
\gridline{\fig{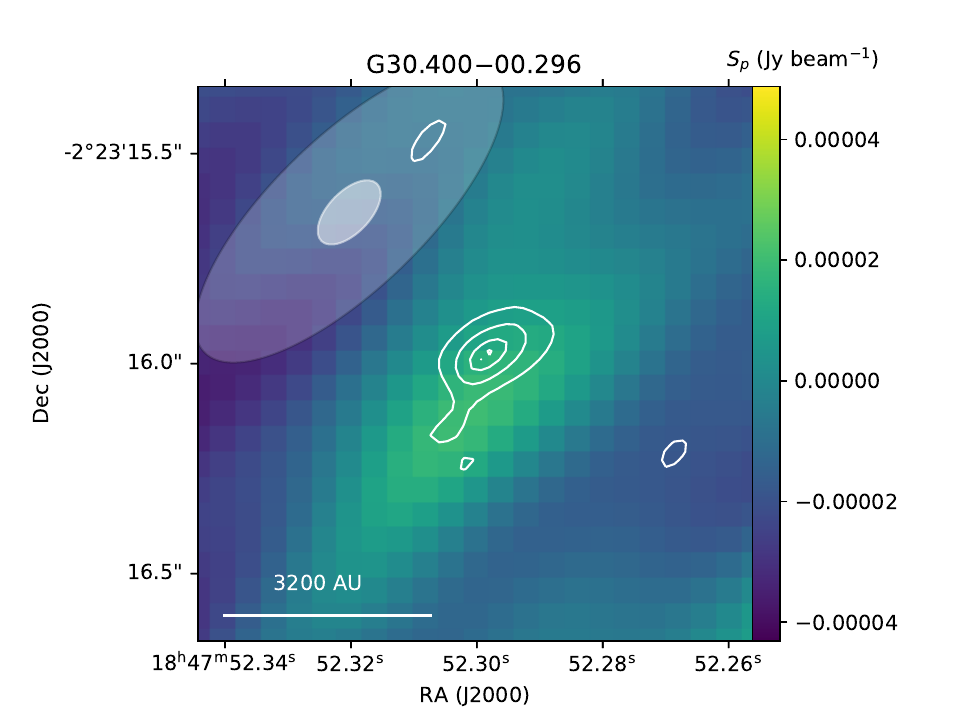}{0.5\textwidth}{(a)}
          \fig{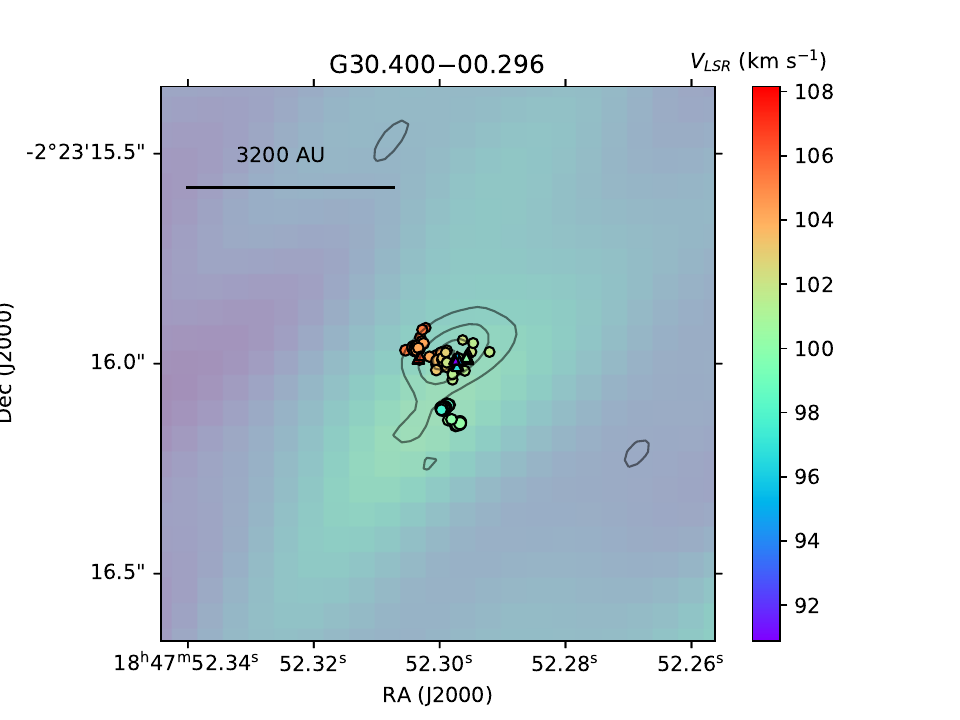}{0.5\textwidth}{(b)}
          }
\gridline{\fig{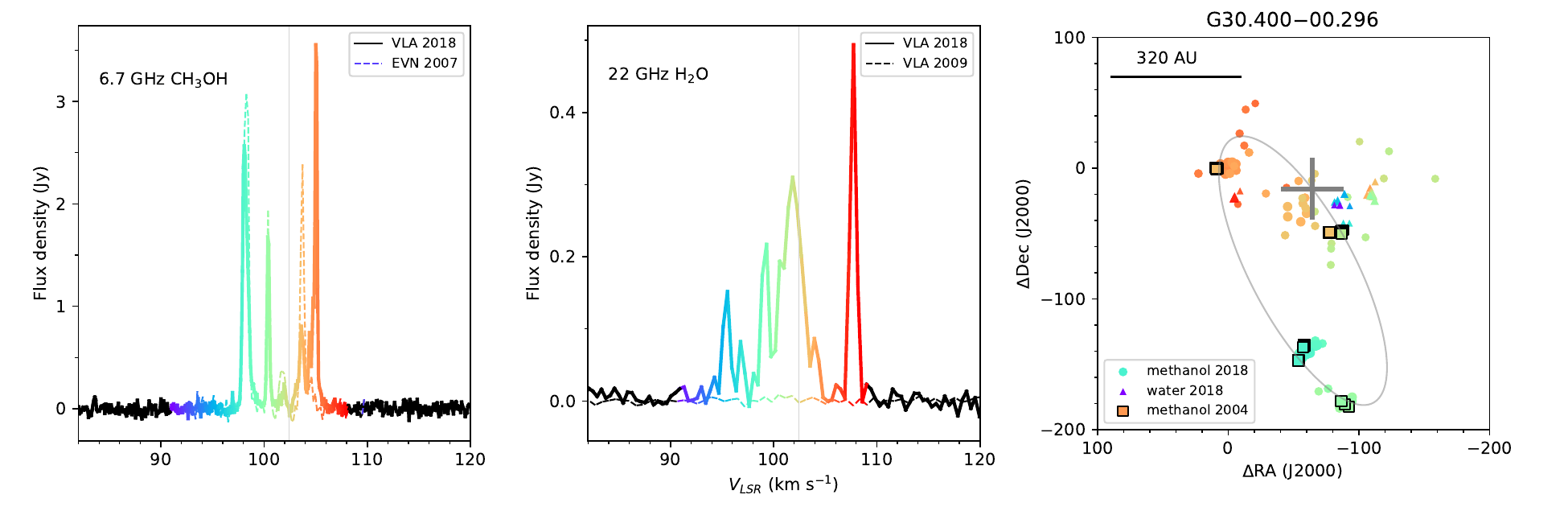}{1\textwidth}{(c)}
          }
\caption{Similar as Figure \ref{fig:G23p207} but for G30.400$-$00.296: (a) The contours of the K-band emission correspond to 3$\sigma$, 7$\sigma$, 11$\sigma$, and 13$\sigma$. 
\label{fig:G30p400}}
\end{figure*}

\begin{figure*}
\gridline{\fig{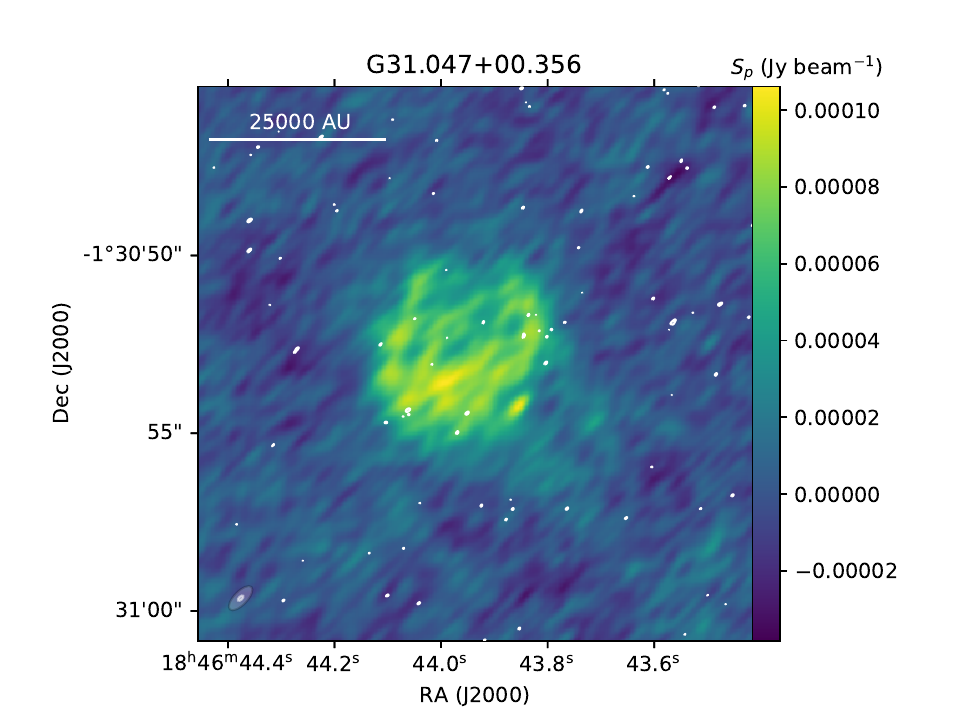}{0.5\textwidth}{(a)}
          \fig{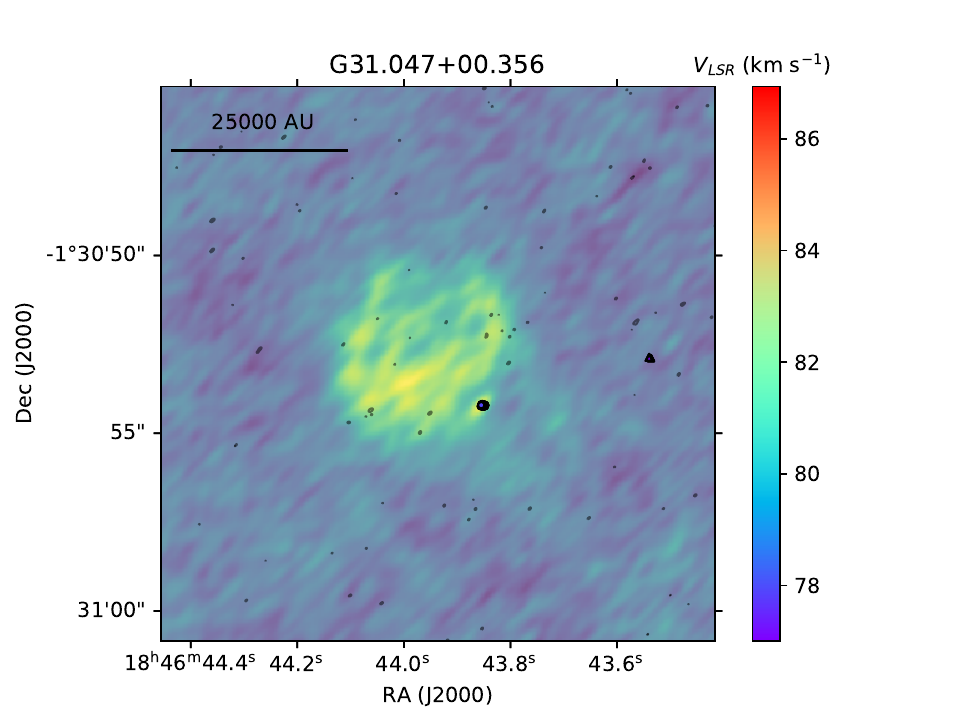}{0.5\textwidth}{(b)}
          }
\gridline{\fig{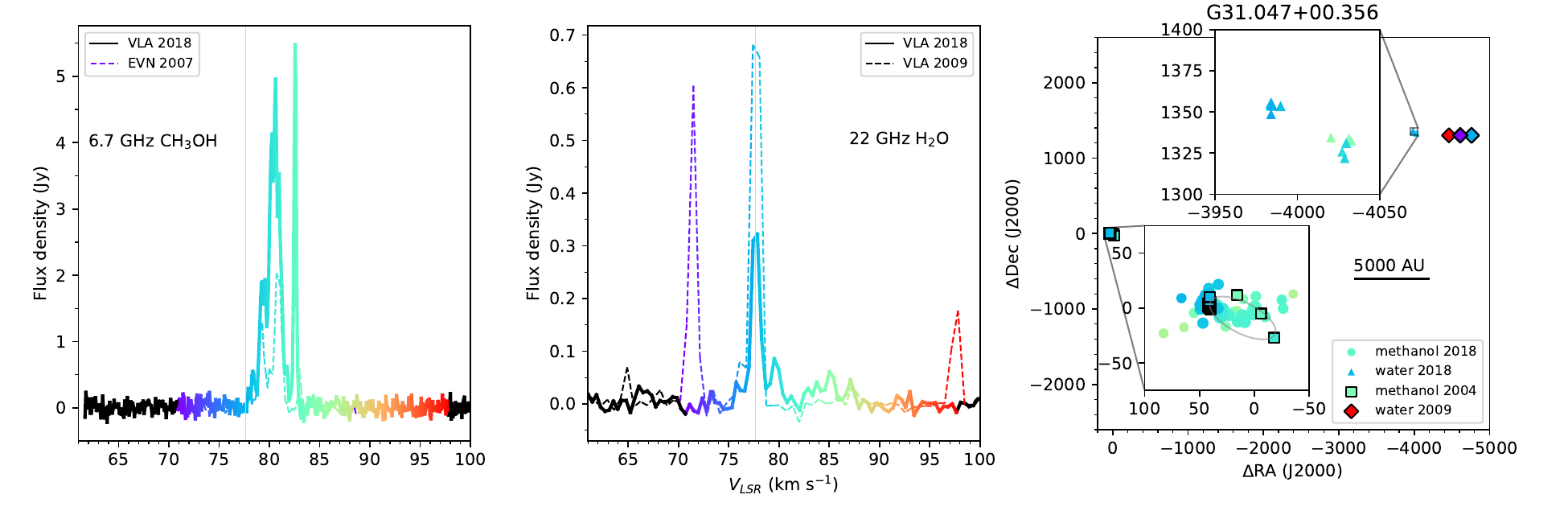}{1\textwidth}{(c)}
          }
\caption{Similar as Figure \ref{fig:G23p207} but for G31.047$+$00.356: (a) The contours of the K-band emission correspond to 3$\sigma$. 
\label{fig:G31p047}}
\end{figure*}

{\it G28.817$+$00.365}. We detected compact radio continua at both bands, their peaks coincided within 0\farcs11 (Figure \ref{fig:G28p817}). This is the case where the 8.4~GHz continuum was reported earlier by \citet{bartkiewicz2009} with a peak flux density of 0.81~Jy~b$^{-1}$. The methanol masers coincide with the SE of the K-band emission and the central part of C-band emission. The methanol maser VLA distribution and spectrum are in agreement with the EVN results \citep{bartkiewicz2009}. However, as noticed in the Torun monitoring program, a systematic increase in the flux density has been seen for all features since 2018 (Figure \ref{fig:monitoring} in the Appendix). Water maser spots are distributed westward over a size of 3". That is comparable with previous observations where the methanol spots are related to the radio continuum and water masers are spread over 2" on the west \citep{bartkiewicz2011}. The water maser spectrum shows significant variability in the brightness (Figure \ref{fig:G28p817}). 

{\it G30.400$-$00.296}. No detection of radio continuum at C band was achieved with a limit of 36~$\mu$Jy~b$^{-1}$ (3$\sigma$). The K-band observations revealed emission with a peak flux density of 66~$\mu$Jy~beam$^{-1}$ with a tentative (3$\sigma$) elongation to the south (Figure \ref{fig:G30p400}). The methanol and water masers coincide with the peak of this radio-continuum. The blue-shifted methanol maser spots might be related to the extended K-band emission. The methanol masers are distributed over a larger area comparing with the EVN images and we notice variability of the red-shifted features at the LSR velocities from 103.3~km~s$^{-1}$ till 105.3~km~s$^{-1}$ \citep{bartkiewicz2009}. In fact, periodic variabilities of the extreme blue- and red-shifted spectral features are noticed with periods of 220-222~days in the single-dish data (Figure \ref{fig:monitoring} in the Appendix). The water maser spots are observed over the same region of the methanol masers, which is contrary to the previous results \citep{bartkiewicz2011}.

{\it G31.047$+$00.356}. We detected extended C-band emission with an exceptionally complex structure over a region of 4"$\times$4" with no counterpart at the K-band with a limit of 18~$\mu$Jy~b$^{-1}$ (3$\sigma$). The methanol masers are located closely to the radio-continuum region, while water maser spots, similarly as it was reported earlier by \citet{bartkiewicz2011}, are found 5" westward and they very likely belong to different region (Figure \ref{fig:G31p047}). The 6.7~GHz methanol maser emission appeared within the same LSR velocities as it was seen using EVN, while the 22~GHz water maser line shows significantly narrower and less bright emission. The Torun monitoring indicates the velocity drift of the red-shifted (ca.~82.5~km~s$^{-1}$) 6.7~GHz spectral feature (Figure \ref{fig:monitoring} in the Appendix). 

\subsection{Spectral Energy Distribution and Luminosities\label{sec:sed}}
In order to measure spectral indexes ($\alpha$) based on the integrated fluxes at each band (Spectral Energy Distribution, SED) we followed the procedure as in \citet{sanna2018}: new images were produced using: (1) a common {\it uv} range of 40-800~k$\lambda$ at both bands, (2) the same beamsize of 0\farcs30$\times$0\farcs3 and the pixel size of 0\farcs06. They are presented in Figure \ref{fig:sed_images} in the Appendix. Further, we assumed there was not variability of the targets over a maximum of 43 days of the time-interval between C- and K-band observations. The $\alpha$ parameter,  $S_{\nu_1}/S_{\nu_0}=(\nu_1/\nu_0)^{\alpha}$ was derived from the ratio between the C- and K-band fluxes; the fits are presented on Figure \ref{fig:SED} in the Appendix. In a case of G31.047$+$00.356, the C-band data were imaged separately for the lower and upper parts of the band, to obtain "in-band" SED since the K-band measurement was not representative as being not sensitive to the large scales.

We have calculated the centimetre luminosities of targets at 8~GHz to discriminate between unresolved H~{\small II} regions and jets \citep{anglada2018}. We also calculated the bolometric (as in \citealt{sarniak2018}), the isotropic methanol and water maser luminosities (according to the formulas: $L_{\rm 6.7~GHz} [L_\odot] = 6.9129\times 10^{-9} \times D^2 [kpc^2] \times S_{\rm int} [Jy~km~s^{-1}]$ and $L_{\rm 22~GHz} [L_\odot] = 23.04996\times 10^{-9} \times D^2 [kpc^2] \times S_{\rm int} [Jy~km~s^{-1}]$, respectively). These results are listed in Table~\ref{tab:summary}, where we summarize the radio continuum and maser detections and list the $\alpha$ values (columns (2)--(6)). We also give the bolometric (L$_{bol}$), isotropic maser (L$_{CH_3OH}$, L$_{H_2O}$) and radio (L$_{\rm 8GHz}$) luminosities (columns (7)--(10)). In column (11) we present the logarithm of the infrared flux ratio $S_\mathrm{70\mu m}/S_\mathrm{22\mu m}$ that will be used when discussing the evolutionary stage below. The flux densities at 22 and 70~$\mu m$ were taken from WISE (\citealt{wright2010}, \citealt{https://doi.org/10.26131/irsa142}) and Herschel (\citealt{molinari2016}) surveys, respectively.

\section{Discussion\label{sec:discussion}}

The spectral indexes of our targets are positive in all cases except G31.047$+$00.356. This slow rise of the flux density with frequency is interpreted as free-free emission from stellar winds and jets (e.g., \citet{anglada2018} and references therein). Comparing the new results with previous observations, one can clearly notice that more sensitive data were necessary to detect radio-continuum emission in all targets. Previously, only G28.817$+$00.365 showed the 8.4~GHz continuum \citep{bartkiewicz2009} which is in agreement with the newly obtained SED distribution. In this publication, we confirm that our previous observations with a sensitivity of 0.15~mJy~beam$^{-1}$ were not enough to detect weak radio emission related to the ring-like methanol masers. 

In the case of G23.207$-$00.377, the emission at K band is resolved in four distinct continuum peaks within a region of about 2000~AU in size. Their overall spectral index between C and K bands is consistent with that of a stellar wind, when the K-band resolution is tapered to match that of the C band (Figure \ref{fig:SED} in the Appendix). Currently, we have to consider two different scenarios that will be tested in the future. In the first scenario, individual continuum peaks would correspond to separated HMYSOs as we note that the 6.7~GHz methanol maser clusters are separated between VLA-1 and VLA-2 and that is possible that we are observing two distinct HMYSOs. In the second scenario, however, the methanol maser clusters may instead trace the edges of a single accretion disc, as both show clear velocity gradients with the redshifted maser spots associated with VLA-1 and the blueshifted spots with VLA-2. The inferred disc size of $\sim$2000 AU falls within the expected range for accretion discs around massive protostars. In this interpretation, the K-band peaks could correspond to jet knots rather than individual YSOs.

In the case of G23.657$-$00.127, we find that at the VLA higher resolution the K-band emission is resolved out, excluding the existence of compact emission brighter than about 21~$\mu$Jy~beam$^{-1}$. A similar result is found towards G31.047$+$00.356 where the K-band emission is resolved out with respect to the C-band emission, although tracing spatial scales much larger than for G23.657$-$00.127. In the case of G31.047$+$00.356, the extended C-band emission is consistent with a compact H~{\small II} region whose spectral index cannot be reliably determined at the moment (see Figure \ref{fig:SED} in the Appendix); therefore, we list its upper limit in the Table~\ref{tab:summary}. 

Correlation between the bolometric luminosity, methanol and water maser luminosities and the radio continuum luminosity at cm wavelengths is presented in Figure \ref{fig:luminosity} in the Appendix. All six targets are associated with the region of high-luminous objects, i.e. above 1000~L$_\odot$, with spectral classes B3 or earlier \citep[e.g.,][]{anglada2018,sanna2018}. G23.207$-$00.377 and G23.389$+$00.185 are located on the best-fit dependence of L$_{\rm 8~GHz}$ and L$_{\rm bol}$ as reported by \citet{sanna2018} meaning that the emission is not dominated by photo-ionization and further supporting that we are tracing a wind of shocked ionized gas from a young star or a tight cluster of young stars as likely in G23.207$-$00.377. The G31.047$+$00.356 source lies close to the expected luminosity due to Lyman continuum flux and supporting the interpretation of an H~{\small II} region. The target G23.657$-$00.127, showing the most circular maser symmetry, also emits the lowest radio luminosity among the sample, as well as the lowest maser luminosity. This low luminosity associated with both tracers might be related to the evolutionary stage or/and the orientation of the system, the latter due to the outflow direction being aligned with line-of-sight. If this interpretation is correct, faint radio continuum might be due to the outflow cavities being mostly devoid of gas at a late stage of evolution. As one may note in Figure \ref{fig:luminosity} in the Appendix, the radio continuum and maser luminosities are weakly related to each other in our sample. The reason can not be verified with these data.

\begin{deluxetable*}{ccccccccccc}
\tablenum{4}
\tablecaption{Detection of methanol and water maser emission\label{tab:summary}}
\tablewidth{0pt}
\tablehead{
\colhead{Target} & \multicolumn2c{Continuum} & {Spectral Index} &  \multicolumn2c{Maser} & {L$_{bol}$} & {{\bf log}(L$_{CH_3OH}$)} & {{\bf log}(L$_{H_2O}$)} & {{\bf log}(L$_{\rm 8GHz}$)} & log($\frac{S_{70\mu m}}{S_{22\mu m}}$)\\
\colhead{Gll.lll$+$bb.bbb} & C-band & K-band & $\alpha_{6-22GHz}$ & methanol & water & (10$^3$L$_\odot$) & (L$_\odot$) & (L$_\odot$) & (mJy~kpc$^2$) & \\
}
\decimalcolnumbers
\startdata
G23.207$-$00.377 &  Y & Y & 0.93$\pm$0.17 & Y & Y  & 13.7 & $-$5.07 & $-$5.31 & 2.51$^{*}$\\
G23.389$+$00.185 &  Y & Y & 0.58$\pm$0.17 & Y & Y  & 12.1 & $-$5.11 & $-$4.66 &  2.81 & 0.54 \\
G23.657$-$00.127 &  Y & N$^{**}$ & ${\bf <}$0.90$\pm$0.90 & Y & N      & 10.9 & $-$5.49 & ${\bf <}-$7.65 &  0.24 & 0.44 \\
G28.817$+$00.365 &  Y & Y & 0.71$\pm$0.06 & Y & Y  & 5.4  & $-$4.55 & $-$5.33 & 12.47 & 1.60 \\
G30.400$-$00.296 &  N & Y & ${\bf >}$0.70$\pm$0.70 & Y & Y  & $-$6.33 & $-$5.41 & 1.83 & 3.2\\
G31.047$+$00.356 &  Y & N$^{**}$ & {\bf $<-1.14\pm0.02^{***}$} & Y & Y  & 5.2  & $-$6.95 & $-$5.28 &  69.97 & 1.15\\
\enddata
\tablecomments{$^*$ the luminosity representative of the whole core and not of a single YSO, $^{**}$ likely resolved emission, $^{***}$ the extended structure at the C-band does not allow to estimate the reliable flux density}.
\end{deluxetable*}

Our proper motion studies of methanol maser cloudlets for these methanol rings have revealed that radial motions dominate over tangential motions with the velocity vectors aligned with the major axis of the methanol maser distribution \citep{bartkiewicz2024}. Therefore, it is likely that the methanol maser kinematics is tracing the radial expansion of a disc or an envelope. In G23.207$-$00.377 and G23.389$+$00.185 the major axes of such inclined discs/envelopes are roughly perpendicular to the outflow direction as traced by the VLA K-band continuum assuming the scenario with a single young star. The most circular methanol maser distribution in G23.657$-$00.127 also shows expanding motions \citep{bartkiewicz2020}. As we stated above, a thermal jet or a stellar wind directed toward the observer is consistent with the radio continuum images and their spectral indexes. Under these conditions, the exciting source, HMYSO, would be surrounded by the masing cloudlets as proposed by \citet{bartkiewicz2020}, their Scenario~II -- the methanol masers would be expanding each side of the jet axis, and trace a wide angle wind at the base of the protostellar jet similarly as in G23.01$-$0.41 \citep{sanna2010b}. At the moment, we can only speculate that these source properties might be related to a later stage of evolution, when also water masers are quenched. We note that this is contrary for our postulate in \citep{bartkiewicz2011} that methanol maser rings are related to the earliest stages of evolution of the young stars they are associated with, before photo-ionization takes place and before outflows start to hit the circumstellar environment. But a later stage of evolution would be in agreement with a recent conclusion by \citet{ladeyschikov2024} that 22~GHz water masers may precede 6.7~GHz methanol masers in the evolution timeline of star-formation regions. However, we note for completeness that a previous "straw man" evolutionary sequence for masers in high-mass star formation regions proposed by \citet{ellingsen2007} suggests the opposite. \citet{debuizer2012} detected near- and mid-infrared emission both in G23.389$+$00.185 and G23.657$-$00.127 peaking at a position slightly shifted from the centers of the methanol maser rings. Such extended emission, if directly associated with the HMYSO driving the methanol masers, indicates a rather more evolved stage of evolution. In G28.817$+$00.365, the water maser emission distributed along the east-west strongly suggests a jet in that direction with a size of ca.~70~mpc, and there is also a velocity gradient along the north-south direction traced by the 6.7~GHz methanol masers. We note that the K-band continuum is compact, but it shows a tentative elongation towards the north-east direction which might be indicative of a jet that should be inspected further. In G30.400$-$00.296, maser proper motions are only derived tentatively (Bartkiewicz et al. {\it in prep.}), the methanol maser emission distributes nearly perpendicularly to the NW-SE elongation of the K-band emission, the latter possibly tracing a thermal jet. Under this assumption, the maser velocity gradient seen from NE to SW may indicate rotation. Finally, the methanol masers in G31.047$+$00.356 trace a much smaller, arc-like structure where proper motions are mostly perpendicular to the arc \citep{bartkiewicz2024}. It is likely that in this case, the masers are tracing gas driven by the expansion of the  H~{\small II} region.

We verified the evolutionary stage of our targets, as done by \citet{purser2021}, who used the infrared flux ratio $S_\mathrm{70\mu m}/S_\mathrm{22\mu m}$. Reliable estimates of this infrared indicator of HMYSO age were obtained for four objects (Table \ref{tab:summary}), which have values $\leq 1.60$. Comparison with a much larger sample in \citet{purser2021} implies that the HMYSOs associated with the ring-line methanol maser structures are highly evolved. Two objects G23.389$+$00.185 and G23.657$-$00.127 with $L_\mathrm{bol} \geq 1.09\times10^4 L_{\odot}$ and $S_\mathrm{70\mu m}/S_\mathrm{22\mu m} \leq 0.55 $ appear to be the oldest among our sample. In contrast, G28.817$+$00.365, with its jet traced also by the water maser, would be the youngest in this sample. 

\section{Conclusions}
We report on the sensitive, multiband VLA observations towards six HMYSOs related to the ringlike 6.7~GHz methanol masers. We detected radio-continuum in all targets, but in four we detected emission at both bands: 6 and 22~GHz. Careful estimations and analysis of SEDs revealed an existence of the thermal radio jets in five out of six sources. In G23.389$+$00.185, G23.657$-$00.127, G28.817$+$00.365, and G30.400$-$00.296, there is clear evidence of compact thermal jet detections; furthermore, the G23.207$-$00.377 target, where we report the complex radio-continuum emission, we also propose a jet candidate due to the water maser relation. In G31.047$+$00.356, we expect a compact H~{\small II} region based on its C-band extended morphology and resolved emission at the K band. The compact structures/morphologies of 6.7~GHz maser emission that were imaged using the EVN \citep{bartkiewicz2009} does not depend of the evolutionary stage of HMYSO. The most circular ring G23.657$-$00.127 is likely the oldest in this sample and the weak C-band radio-continuum is likely related to the outflow directed toward the observer; however, the opposite can not be ruled out if the orientations of the system plays a role -- the outflow direction being aligned with the line-of-sight. 

In G23.207$-$00.377, G23.389$+$00.185, G23.657$-$00.127, G28.817$+$00.365, and G30.400$-$00.296, the 6.7~GHz methanol maser spots are related to the peaks of radio-continua. With the existing data we are not able to distinguish, if we recovered the resolved emission by the VLBI or captured the emission from closely separated methanol masers forming continuous structures like in G23.207$-$00.377. The water maser transition is variable in general; therefore, the comparison of only two epochs of observations over ca.~14 years is of a little value. However, the significant changes were pointed in five targets expect in G23.657$-$00.127. In the latter case, we again, do not detect any 22~GHz water masers. This can be also related to the evolutionary stage of the HMYSO and/or the special orientation of the system. 

Still, multi-frequency studies at a complementary angular resolution of thermal tracers are needed to explore environment of borning high-mass stars including discs and envelopes.

\begin{acknowledgments}
AB, AK, MSz acknowledge support from the National Science Centre, Poland through grant 2021/43/B/ST9/02008. We thank the National Radio Astronomy Observatory of the USA for the opportunity to observe with the VLA and the NRAO New Mexico staff for their assistance in carrying out the observations. The National Radio Astronomy Observatory is a facility of the National Science Foundation operated under cooperative agreement by Associated Universities, Inc.
\end{acknowledgments}



%
\facilities{VLA, NCU:RT4}

\software{CASA \citep{McMullin07}, Python 3 (\url{https://www.python.org/})}


\appendix
\section{Additional figures\label{appendix}}
\begin{figure*}
\gridline{\fig{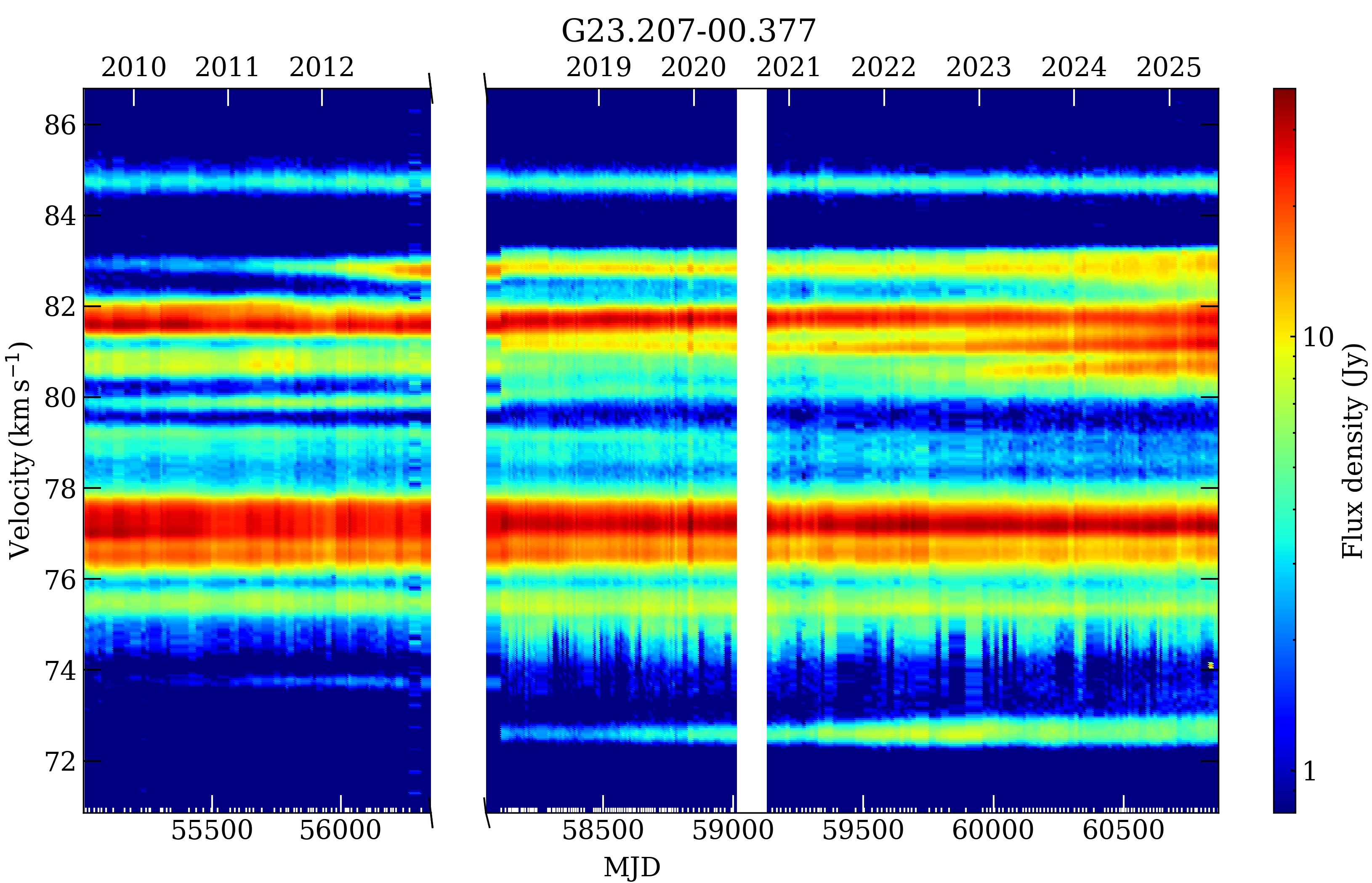}{0.5\textwidth}{(a)}\hspace{0.25cm}
         \fig{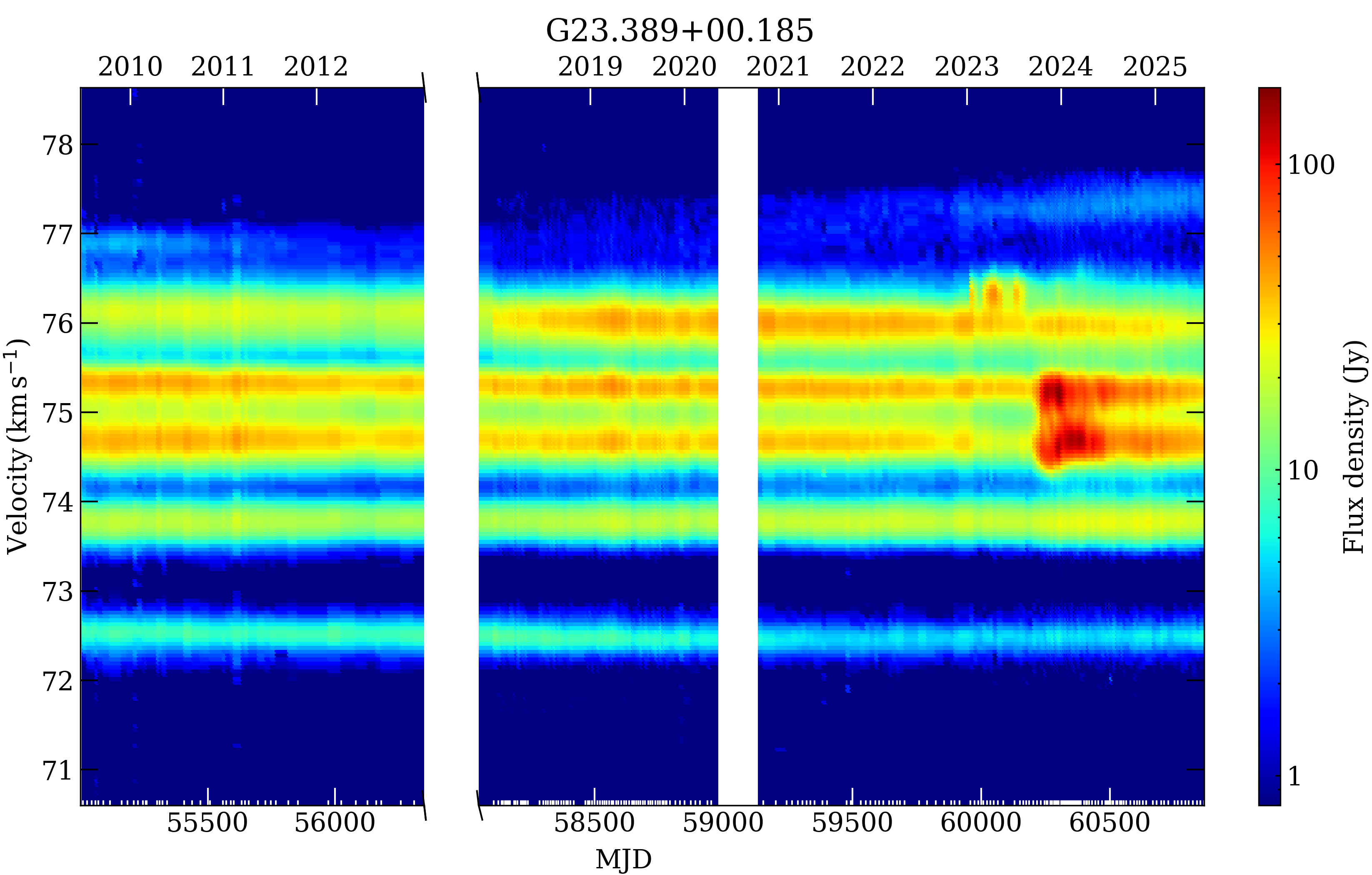}{0.5\textwidth}{(b)}
         }
\gridline{\fig{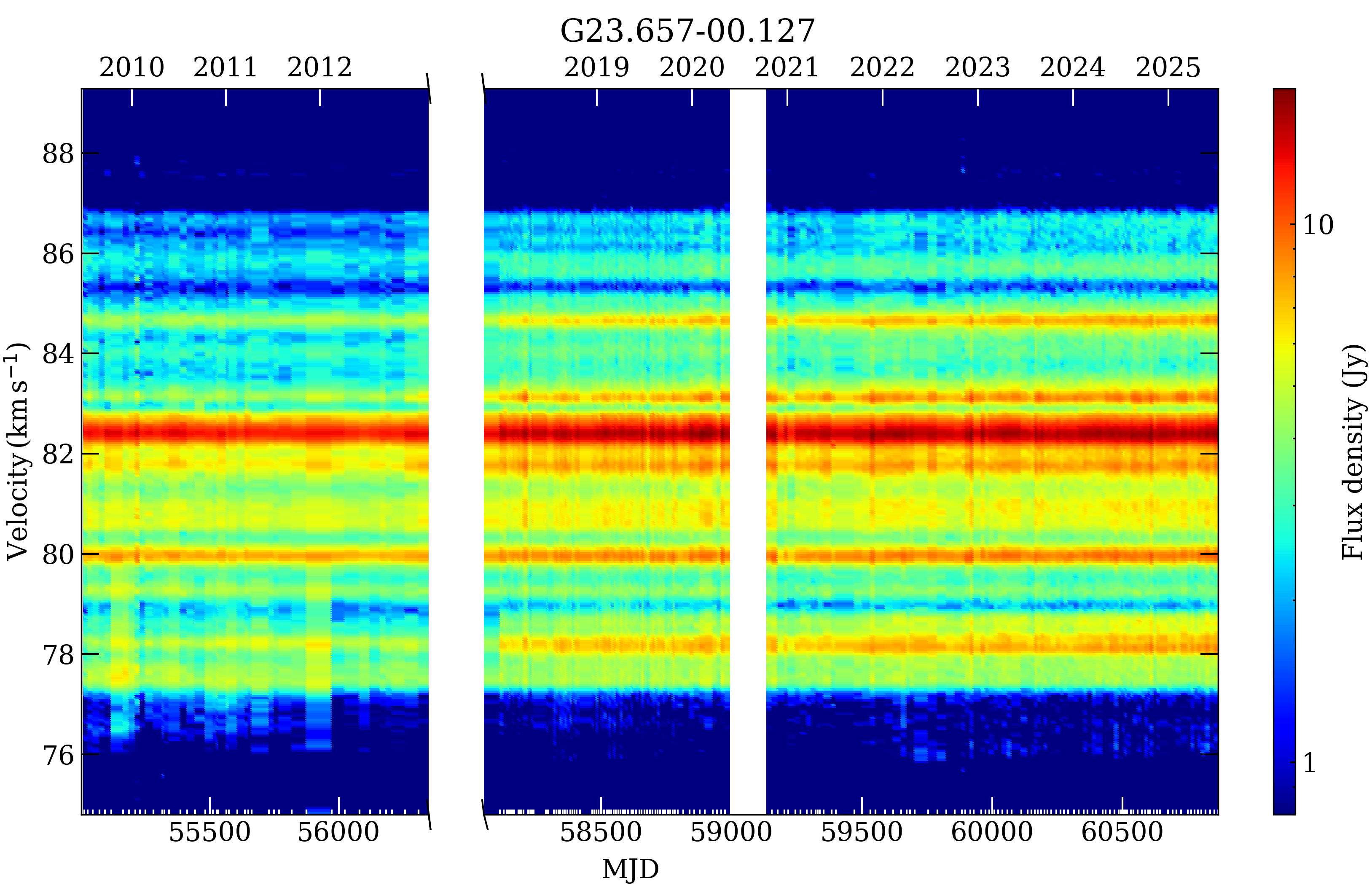}{0.5\textwidth}{(c)}\hspace{0.25cm}
          \fig{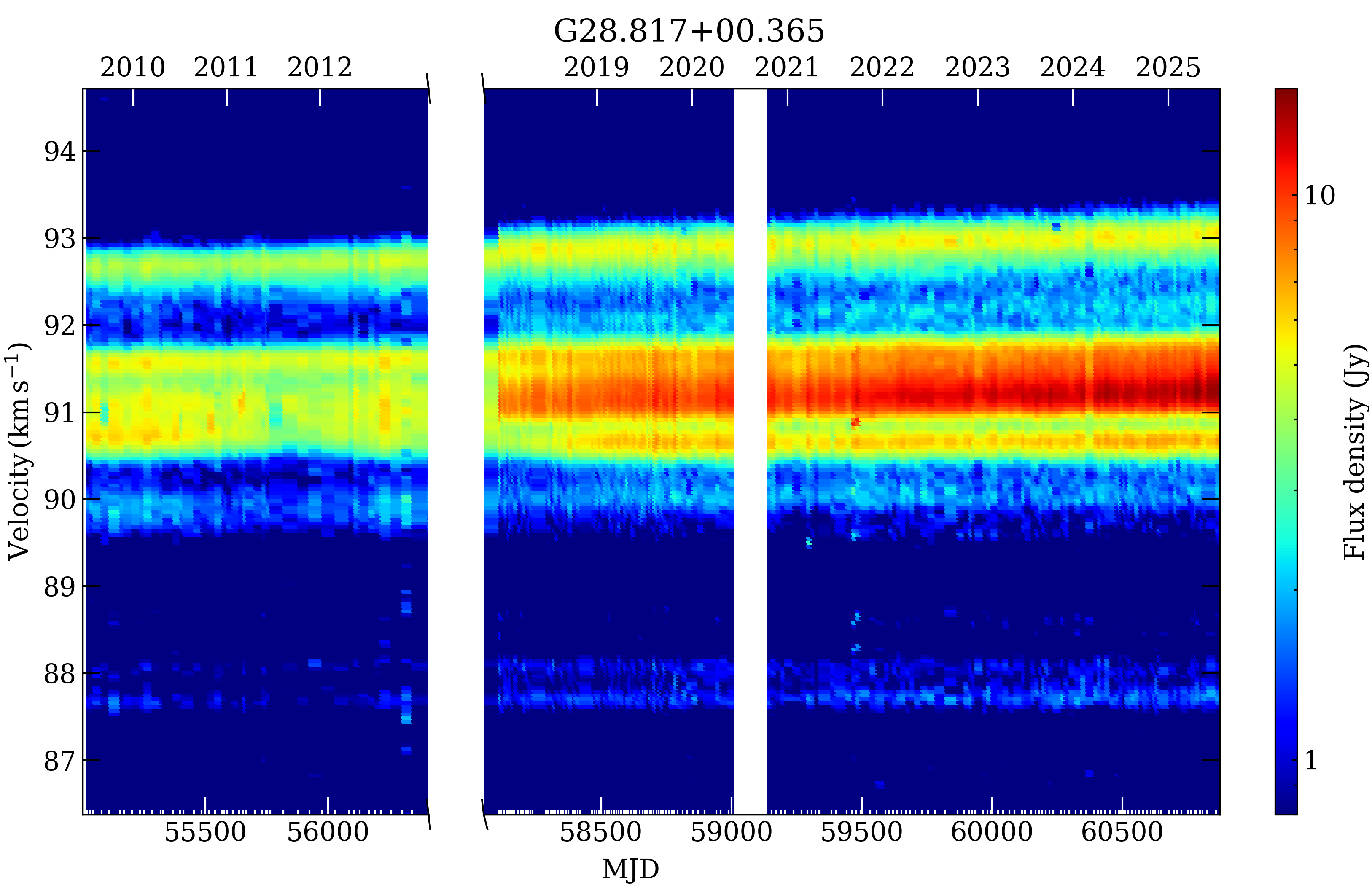}{0.5\textwidth}{(d)}
          }
\gridline{\fig{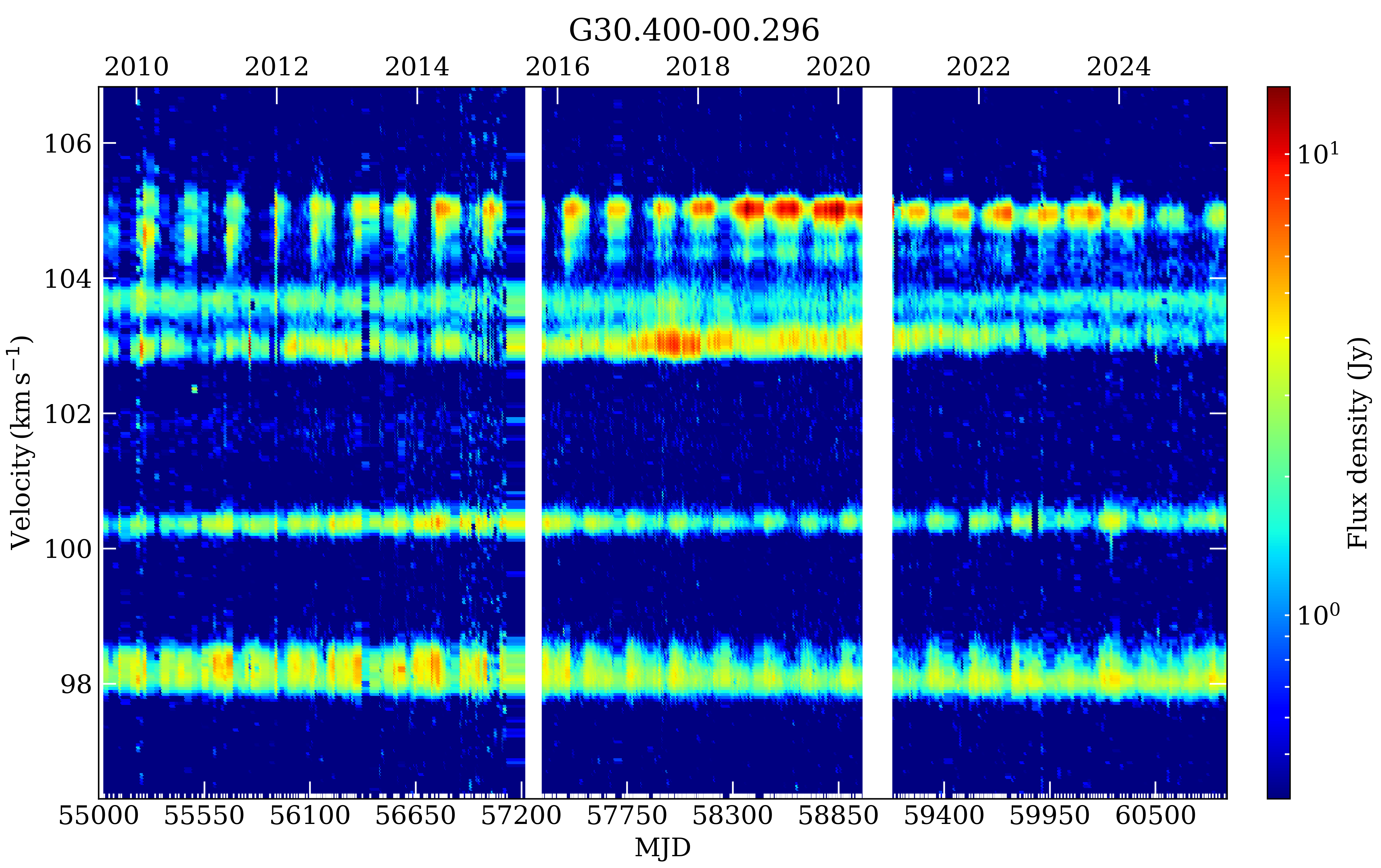}{0.5\textwidth}{(e)}\hspace{0.25cm}
          \fig{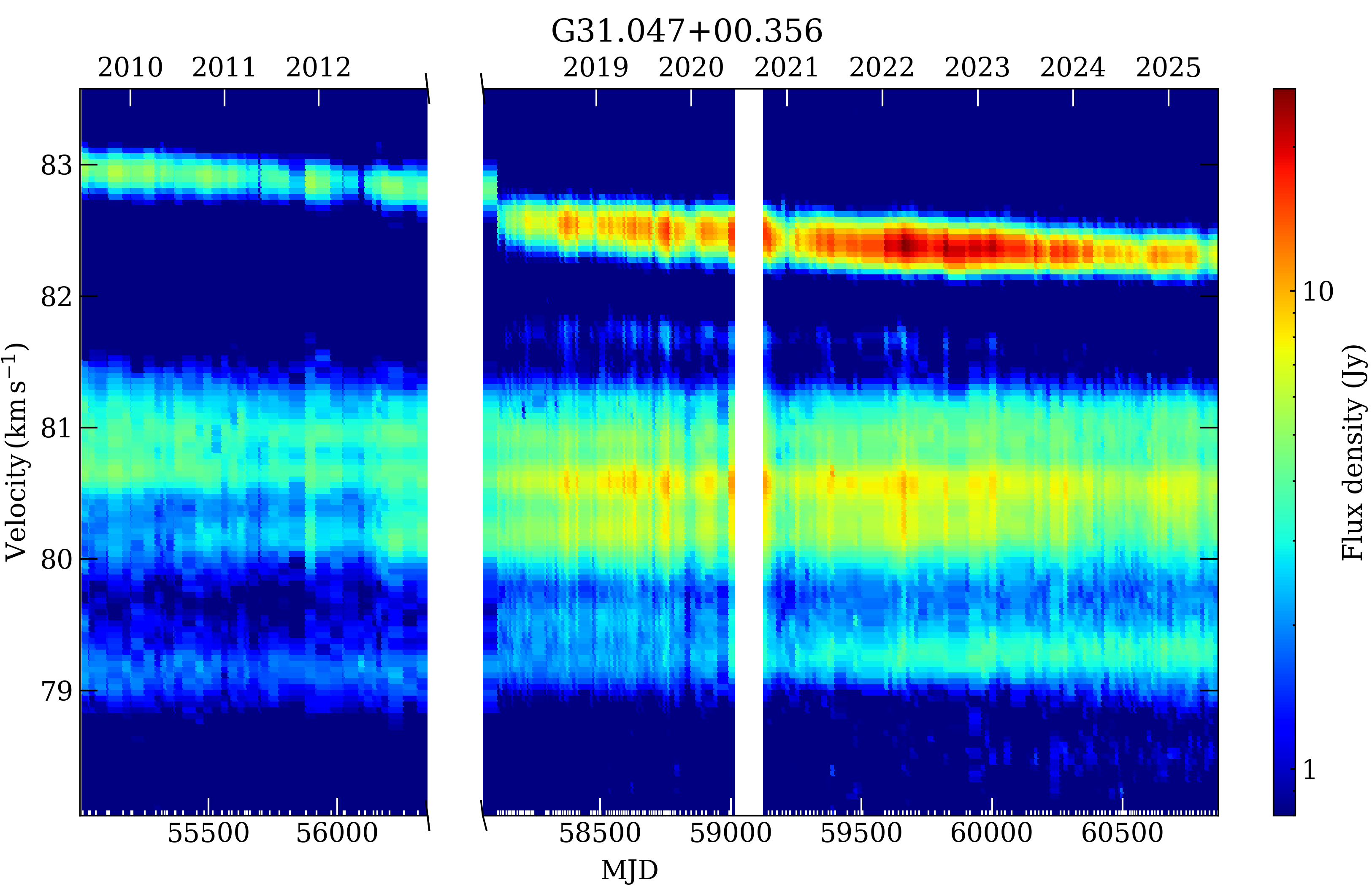}{0.5\textwidth}{(f)}}

\caption{The dynamic spectra of the variability of the 6.7~GHz methanol maser lines towards the targets as monitored using the 32~m Torun dish. The epochs corresponding to each observation are marked by the minor ticks on the lower x-axes of all figures.
\label{fig:monitoring}}
\end{figure*}

\begin{figure*}
\gridline{\fig{bartkiewicz_continuum_sed_g23.207.pdf}{0.5\textwidth}{(a)}
          \fig{bartkiewicz_continuum_sed_g23.389.pdf}{0.5\textwidth}{(b)}
          }
\gridline{\fig{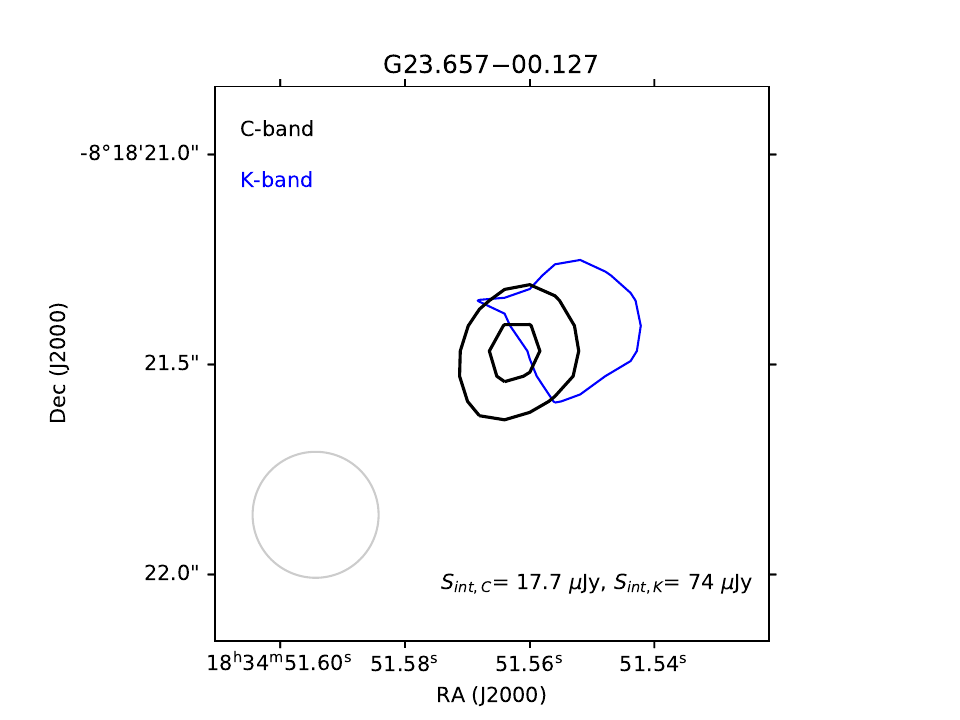}{0.5\textwidth}{(c)}
          \fig{bartkiewicz_continuum_sed_g28.817.pdf}{0.5\textwidth}{(d)}
          }
\gridline{\fig{bartkiewicz_continuum_sed_g30.400.pdf}{0.5\textwidth}{(e)}
          \fig{bartkiewicz_continuum_sed_g31.047.pdf}{0.5\textwidth}{(f)}
          }
\caption{The images of radio-continuum emission at 6 and 22~GHz created homogeneously for the SED analysis. The contours in panels (a), (b), (d)-(f) are the same as in Figures \ref{fig:G23p207}-\ref{fig:G31p047}, in a case of panel (c), K-band data 1$\sigma$ is marked. A synthesized beam size of 0\farcs30$\times$0\farcs30 (marked by a gray circle) and a pixel size of 0\farcs06 were used for all images at the C and K bands. The integrated fluxes at both transitions are listed.
\label{fig:sed_images}}
\end{figure*}

\begin{figure*}
\fig{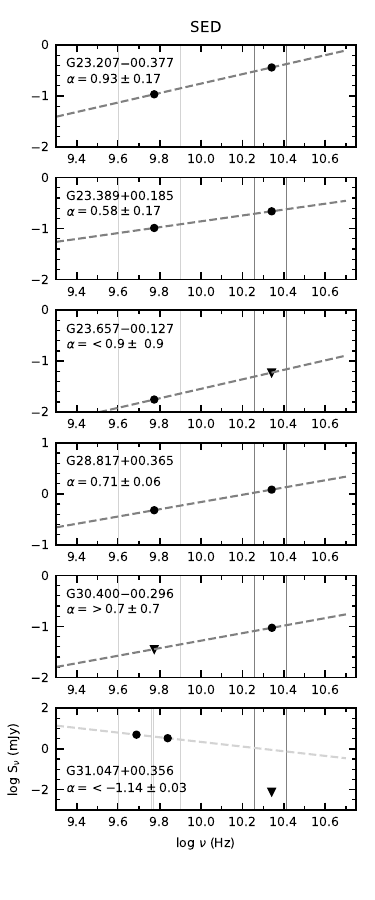}{0.5\textwidth}{}
\caption{The SED plots with the spectral index $\alpha$ fitted and marked by the dashed lines. The vertical gray lines correspond to C- and K-band ranges. The lower limits of the flux densities are marked by the downward-pointing filled triangles.
\label{fig:SED}}
\end{figure*}

\begin{figure*}
\fig{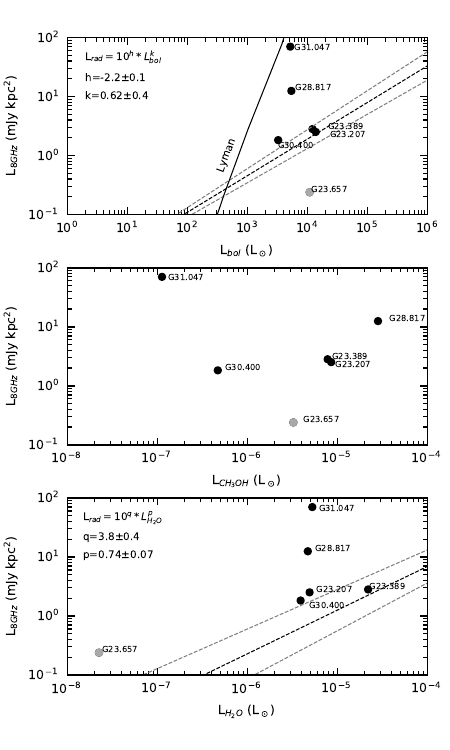}{0.5\textwidth}{}
\caption{Dependences of radio luminosity of the methanol ringlike sources on the bolometric (top), methanol (middle), and water maser luminosities (bottom). The gray dot corresponds to the most circular target G23.657$-$00.127. The dotted lines refer to fits by \citet{sanna2018} (their Figure 5) and their parameters are also listed in the left top corners.
\label{fig:luminosity}}
\end{figure*}

This appendix contains figures relating to the 6.7~GHz methanol maser line variation monitoring program (Section \ref{sec:torunmon}), the method for determining the spectral energy distribution (SED) and verifying the luminosity dependencies
(Section \ref{sec:sed}). In Figure \ref{fig:monitoring} we present the dynamic spectra of
the variability of the 6.7~GHz methanol maser lines as monitored using the 32 m Torun dish. In Figure \ref{fig:sed_images} we show the images of radio-continuum emission at 6 and 22~GHz created homogeneously for the SED analysis, as well as the spectral
indexes $\alpha$ fits in Figure \ref{fig:SED}. Dependences of radio luminosity of
the methanol ringlike sources on the bolometric, methanol, and water maser luminosities are presented in Figure \ref{fig:luminosity}.

\bibliography{bartkiewicz}{}

@ARTICLE{anglada2018,
       author = {{Anglada}, Guillem and {Rodr{\'\i}guez}, Luis F. and {Carrasco-Gonz{\'a}lez}, Carlos},
        title = "{Radio jets from young stellar objects}",
      journal = {\aapr},
         year = 2018,
        month = jun,
       volume = {26},
       number = {1},
          eid = {3},
        pages = {3},
          doi = {10.1007/s00159-018-0107-z},
archivePrefix = {arXiv},
       eprint = {1806.06444},
 primaryClass = {astro-ph.SR},
       adsurl = {https://ui.adsabs.harvard.edu/abs/2018A&ARv..26....3A}
}

@ARTICLE{bartkiewicz2024,
       author = {{Bartkiewicz}, A. and {Sanna}, A. and {Szymczak}, M. and {Moscadelli}, L. and {van Langevelde}, H.~J. and {Wolak}, P. and {Kobak}, A. and {Durjasz}, M.},
        title = "{Proper motion study of the 6.7 GHz methanol maser rings. I. A sample of sources with little variation}",
      journal = {\aap},
         year = 2024,
        month = jun,
       volume = {686},
          eid = {A275},
        pages = {A275},
          doi = {10.1051/0004-6361/202449491},
archivePrefix = {arXiv},
       eprint = {2404.07333},
 primaryClass = {astro-ph.GA},
       adsurl = {https://ui.adsabs.harvard.edu/abs/2024A&A...686A.275B}
}

@ARTICLE{bartkiewicz2020,
       author = {{Bartkiewicz}, A. and {Sanna}, A. and {Szymczak}, M. and {Moscadelli}, L. and {van Langevelde}, H.~J. and {Wolak}, P.},
        title = "{The nature of the methanol maser ring G23.657-00.127. II. Expansion of the maser structure}",
      journal = {\aap},
         year = 2020,
        month = may,
       volume = {637},
          eid = {A15},
        pages = {A15},
          doi = {10.1051/0004-6361/202037562},
archivePrefix = {arXiv},
       eprint = {2004.00916},
 primaryClass = {astro-ph.GA},
       adsurl = {https://ui.adsabs.harvard.edu/abs/2020A&A...637A..15B}
}

@ARTICLE{bartkiewicz2016,
       author = {{Bartkiewicz}, A. and {Szymczak}, M. and {van Langevelde}, H.~J.},
        title = "{European VLBI Network imaging of 6.7 GHz methanol masers}",
      journal = {\aap},
         year = 2016,
        month = mar,
       volume = {587},
          eid = {A104},
        pages = {A104},
          doi = {10.1051/0004-6361/201527541},
archivePrefix = {arXiv},
       eprint = {1601.03197},
 primaryClass = {astro-ph.GA},
       adsurl = {https://ui.adsabs.harvard.edu/abs/2016A&A...587A.104B}
}

@ARTICLE{bartkiewicz2011,
       author = {{Bartkiewicz}, A. and {Szymczak}, M. and {Pihlstr{\"o}m}, Y.~M. and {van Langevelde}, H.~J. and {Brunthaler}, A. and {Reid}, M.~J.},
        title = "{VLA observations of water masers towards 6.7 GHz methanol maser sources}",
      journal = {\aap},
         year = 2011,
        month = jan,
       volume = {525},
          eid = {A120},
        pages = {A120},
          doi = {10.1051/0004-6361/201015235},
archivePrefix = {arXiv},
       eprint = {1009.2334},
 primaryClass = {astro-ph.GA},
       adsurl = {https://ui.adsabs.harvard.edu/abs/2011A&A...525A.120B}
}

@ARTICLE{bartkiewicz2009,
       author = {{Bartkiewicz}, A. and {Szymczak}, M. and {van Langevelde}, H.~J. and {Richards}, A.~M.~S. and {Pihlstr{\"o}m}, Y.~M.},
        title = "{The diversity of methanol maser morphologies from VLBI observations}",
      journal = {\aap},
         year = 2009,
        month = jul,
       volume = {502},
       number = {1},
        pages = {155-173},
          doi = {10.1051/0004-6361/200912250},
archivePrefix = {arXiv},
       eprint = {0905.3469},
 primaryClass = {astro-ph.GA},
       adsurl = {https://ui.adsabs.harvard.edu/abs/2009A&A...502..155B}
}

@ARTICLE{bartkiewicz2008,
       author = {{Bartkiewicz}, A. and {Brunthaler}, A. and {Szymczak}, M. and {van Langevelde}, H.~J. and {Reid}, M.~J.},
        title = "{The nature of the methanol maser ring G23.657-00.127. I. The distance through trigonometric parallax measurements}",
      journal = {\aap},
         year = 2008,
        month = nov,
       volume = {490},
       number = {2},
        pages = {787-792},
          doi = {10.1051/0004-6361:200810470},
archivePrefix = {arXiv},
       eprint = {0809.1948},
 primaryClass = {astro-ph},
       adsurl = {https://ui.adsabs.harvard.edu/abs/2008A&A...490..787B}
}

@ARTICLE{bartkiewicz2005,
       author = {{Bartkiewicz}, A. and {Szymczak}, M. and {van Langevelde}, H.~J.},
        title = "{Ring shaped 6.7 GHz methanol maser emission around a young high-mass star}",
      journal = {\aap},
         year = 2005,
        month = nov,
       volume = {442},
       number = {3},
        pages = {L61-L64},
          doi = {10.1051/0004-6361:200500190},
archivePrefix = {arXiv},
       eprint = {astro-ph/0509578},
 primaryClass = {astro-ph},
       adsurl = {https://ui.adsabs.harvard.edu/abs/2005A&A...442L..61B}
}

@ARTICLE{bayandina2022,
       author = {{Bayandina}, O.~S. and {Brogan}, C.~L. and {Burns}, R.~A. and {Chen}, X. and {Hunter}, T.~R. and {Kurtz}, S.~E. and {MacLeod}, G.~C. and {Sobolev}, A.~M. and {Sugiyama}, K. and {Val'tts}, I.~E. and {Yonekura}, Y.},
        title = "{A Multitransition Methanol Maser Study of the Accretion Burst Source G358.93-0.03-MM1}",
      journal = {\aj},
         year = 2022,
        month = feb,
       volume = {163},
       number = {2},
          eid = {83},
        pages = {83},
          doi = {10.3847/1538-3881/ac42d2},
archivePrefix = {arXiv},
       eprint = {2201.12075},
 primaryClass = {astro-ph.GA},
       adsurl = {https://ui.adsabs.harvard.edu/abs/2022AJ....163...83B}
}

@ARTICLE{breen2013,
       author = {{Breen}, S.~L. and {Ellingsen}, S.~P. and {Contreras}, Y. and {Green}, J.~A. and {Caswell}, J.~L. and {Stevens}, J.~B. and {Dawson}, J.~R. and {Voronkov}, M.~A.},
        title = "{Confirmation of the exclusive association between 6.7-GHz methanol masers and high-mass star formation regions}",
      journal = {\mnras},
         year = 2013,
        month = oct,
       volume = {435},
       number = {1},
        pages = {524-530},
          doi = {10.1093/mnras/stt1315},
archivePrefix = {arXiv},
       eprint = {1307.4453},
 primaryClass = {astro-ph.GA},
       adsurl = {https://ui.adsabs.harvard.edu/abs/2013MNRAS.435..524B}
}

@ARTICLE{chibueze2017,
       author = {{Chibueze}, James O. and {Csengeri}, Timea and {Tatematsu}, Ken'ichi and {Hasegawa}, Tetsuo and {Iguchi}, Satoru and {Alhassan}, Jibrin A. and {Higuchi}, Aya E. and {Bontemps}, Sylvain and {Menten}, Karl M.},
        title = "{Class II 6.7 GHz Methanol Maser Association with Young Massive Cores Revealed by ALMA}",
      journal = {\apj},
         year = 2017,
        month = feb,
       volume = {836},
       number = {1},
          eid = {59},
        pages = {59},
          doi = {10.3847/1538-4357/836/1/59},
       adsurl = {https://ui.adsabs.harvard.edu/abs/2017ApJ...836...59C}
}

@ARTICLE{debuizer2012,
       author = {{De Buizer}, James M. and {Bartkiewicz}, Anna and {Szymczak}, Marian},
        title = "{Testing the Hypothesis that Methanol Maser Rings Trace Circumstellar Disks: High-resolution Near-infrared and Mid-infrared Imaging}",
      journal = {\apj},
         year = 2012,
        month = aug,
       volume = {754},
       number = {2},
          eid = {149},
        pages = {149},
          doi = {10.1088/0004-637X/754/2/149},
archivePrefix = {arXiv},
       eprint = {1206.0055},
 primaryClass = {astro-ph.SR},
       adsurl = {https://ui.adsabs.harvard.edu/abs/2012ApJ...754..149D}
}

@INPROCEEDINGS{ellingsen2007,
       author = {{Ellingsen}, S.~P. and {Voronkov}, M.~A. and {Cragg}, D.~M. and {Sobolev}, A.~M. and {Breen}, S.~L. and {Godfrey}, P.~D.},
        title = "{Investigating high-mass star formation through maser surveys}",
    booktitle = {Astrophysical Masers and their Environments},
         year = 2007,
       editor = {{Chapman}, Jessica M. and {Baan}, Willem A.},
       series = {IAU Symposium},
       volume = {242},
        month = mar,
        pages = {213-217},
          doi = {10.1017/S1743921307012999},
archivePrefix = {arXiv},
       eprint = {0705.2906},
 primaryClass = {astro-ph},
       adsurl = {https://ui.adsabs.harvard.edu/abs/2007IAUS..242..213E}
}

@ARTICLE{hofner2011,
       author = {{Hofner}, P. and {Kurtz}, S. and {Ellingsen}, S.~P. and {Menten}, K.~M. and {Wyrowski}, F. and {Araya}, E.~D. and {Loinard}, L. and {Rodr{\'\i}guez}, L.~F. and {Cesaroni}, R.},
        title = "{Expanded Very Large Array Continuum Observations toward Hot Molecular Core Candidates}",
      journal = {\apjl},
         year = 2011,
        month = sep,
       volume = {739},
       number = {1},
          eid = {L17},
        pages = {L17},
          doi = {10.1088/2041-8205/739/1/L17},
       adsurl = {https://ui.adsabs.harvard.edu/abs/2011ApJ...739L..17H}
}

@ARTICLE{fitzgibbon99,
       author = {{Fitzgibbon}, A. and {Pilu}, M. and {Fisher}, R.~B.},
        title = "{Direct least square fitting of ellipses}",
      journal = {IEEE Trans. Pattern Anal. Mach. Intell.},
         year = 1999,
       volume = {21},
        pages = {476},
          url = {https://ieeexplore.ieee.org/document/765658},
}

@ARTICLE{hu2016,
       author = {{Hu}, B. and {Menten}, K.~M. and {Wu}, Y. and {Bartkiewicz}, A. and {Rygl}, K. and {Reid}, M.~J. and {Urquhart}, J.~S. and {Zheng}, X.},
        title = "{On the Relationship of UC HII Regions and Class II Methanol Masers. I. Source Catalogs}",
      journal = {\apj},
         year = 2016,
        month = dec,
       volume = {833},
       number = {1},
          eid = {18},
        pages = {18},
          doi = {10.3847/0004-637X/833/1/18},
archivePrefix = {arXiv},
       eprint = {1609.03280},
 primaryClass = {astro-ph.GA},
       adsurl = {https://ui.adsabs.harvard.edu/abs/2016ApJ...833...18H}
}

@misc{https://doi.org/10.26131/irsa142,
  doi = {10.26131/IRSA142},
  url = {https://catcopy.ipac.caltech.edu/dois/doi.php?id=10.26131/IRSA142},
  author = {{WISE Team}},
  title = {WISE All-Sky Source Catalog},
  publisher = {IPAC},
  year = {2020}
}

@INPROCEEDINGS{ladeyschikov2024,
       author = {{Ladeyschikov}, Dmitry},
        title = "{Early Star Formation Traced by Water Masers}",
    booktitle = {Cosmic Masers: Proper Motion Toward the Next-Generation Large Projects},
         year = 2024,
       editor = {{Hirota}, Tomoya and {Imai}, Hiroshi and {Menten}, Karl and {Pihlstr{\"o}m}, Ylva},
       series = {IAU Symposium},
       volume = {380},
        month = jan,
        pages = {230-231},
          doi = {10.1017/S1743921323003290},
       adsurl = {https://ui.adsabs.harvard.edu/abs/2024IAUS..380..230L}
}

@ARTICLE{menten1991,
       author = {{Menten}, Karl M.},
        title = "{The Discovery of a New, Very Strong, and Widespread Interstellar Methanol Maser Line}",
      journal = {\apjl},
         year = 1991,
        month = oct,
       volume = {380},
        pages = {L75},
          doi = {10.1086/186177},
       adsurl = {https://ui.adsabs.harvard.edu/abs/1991ApJ...380L..75M}
}

@ARTICLE{moscadelli2016,
       author = {{Moscadelli}, L. and {S{\'a}nchez-Monge}, {\'A}. and {Goddi}, C. and {Li}, J.~J. and {Sanna}, A. and {Cesaroni}, R. and {Pestalozzi}, M. and {Molinari}, S. and {Reid}, M.~J.},
        title = "{Outflow structure within 1000 au of high-mass YSOs. I. First results from a combined study of maser and radio continuum emission}",
      journal = {\aap},
         year = 2016,
        month = jan,
       volume = {585},
          eid = {A71},
        pages = {A71},
          doi = {10.1051/0004-6361/201526238},
       adsurl = {https://ui.adsabs.harvard.edu/abs/2016A&A...585A..71M}
}

@ARTICLE{purser2021,
       author = {{Purser}, S.~J.~D. and {Lumsden}, S.~L. and {Hoare}, M.~G. and {Kurtz}, S.},
        title = "{A Galactic survey of radio jets from massive protostars}",
      journal = {\mnras},
         year = 2021,
        month = jun,
       volume = {504},
       number = {1},
        pages = {338-355},
          doi = {10.1093/mnras/stab747},
archivePrefix = {arXiv},
       eprint = {2103.08990},
 primaryClass = {astro-ph.GA},
       adsurl = {https://ui.adsabs.harvard.edu/abs/2021MNRAS.504..338P}
}

@ARTICLE{sanna2015,
       author = {{Sanna}, A. and {Surcis}, G. and {Moscadelli}, L. and {Cesaroni}, R. and {Goddi}, C. and {Vlemmings}, W.~H.~T. and {Caratti o Garatti}, A.},
        title = "{Velocity and magnetic fields within 1000 AU of a massive YSO}",
      journal = {\aap},
         year = 2015,
        month = nov,
       volume = {583},
          eid = {L3},
        pages = {L3},
          doi = {10.1051/0004-6361/201526806},
archivePrefix = {arXiv},
       eprint = {1509.05428},
 primaryClass = {astro-ph.SR},
       adsurl = {https://ui.adsabs.harvard.edu/abs/2015A&A...583L...3S}
}

@ARTICLE{sanna2018,
       author = {{Sanna}, A. and {Moscadelli}, L. and {Goddi}, C. and {Krishnan}, V. and {Massi}, F.},
        title = "{Protostellar Outflows at the EarliesT Stages (POETS). I. Radio thermal jets at high resolution nearby H$_{2}$O maser sources}",
      journal = {\aap},
         year = 2018,
        month = nov,
       volume = {619},
          eid = {A107},
        pages = {A107},
          doi = {10.1051/0004-6361/201833573},
archivePrefix = {arXiv},
       eprint = {1807.06680},
 primaryClass = {astro-ph.SR},
       adsurl = {https://ui.adsabs.harvard.edu/abs/2018A&A...619A.107S}
}

@INPROCEEDINGS{sarniak2018,
       author = {{Sarniak}, R. and {Szymczak}, M. and {Bartkiewicz}, A.},
        title = "{Statistical analysis of the physical properties of the 6.7 GHz methanol maser features based on VLBI data}",
    booktitle = {Astrophysical Masers: Unlocking the Mysteries of the Universe},
         year = 2018,
       editor = {{Tarchi}, A. and {Reid}, M.~J. and {Castangia}, P.},
       series = {IAU Symposium},
       volume = {336},
        month = aug,
        pages = {321-322},
          doi = {10.1017/S1743921317009425},
       adsurl = {https://ui.adsabs.harvard.edu/abs/2018IAUS..336..321S}
}

@ARTICLE{reid2019,
       author = {{Reid}, M.~J. and {Menten}, K.~M. and {Brunthaler}, A. and {Zheng}, X.~W. and {Dame}, T.~M. and {Xu}, Y. and {Li}, J. and {Sakai}, N. and {Wu}, Y. and {Immer}, K. and {Zhang}, B. and {Sanna}, A. and {Moscadelli}, L. and {Rygl}, K.~L.~J. and {Bartkiewicz}, A. and {Hu}, B. and {Quiroga-Nu{\~n}ez}, L.~H. and {van Langevelde}, H.~J.},
        title = "{Trigonometric Parallaxes of High-mass Star-forming Regions: Our View of the Milky Way}",
      journal = {\apj},
         year = 2019,
        month = nov,
       volume = {885},
       number = {2},
          eid = {131},
        pages = {131},
          doi = {10.3847/1538-4357/ab4a11},
archivePrefix = {arXiv},
       eprint = {1910.03357},
 primaryClass = {astro-ph.GA},
       adsurl = {https://ui.adsabs.harvard.edu/abs/2019ApJ...885..131R}
}

@ARTICLE{szymczak2007,
       author = {{Szymczak}, M. and {Bartkiewicz}, A. and {Richards}, A.~M.~S.},
        title = "{A multi-transition molecular line study of candidate massive young stellar objects associated with methanol masers}",
      journal = {\aap},
         year = 2007,
        month = jun,
       volume = {468},
       number = {2},
        pages = {617-625},
          doi = {10.1051/0004-6361:20077289},
archivePrefix = {arXiv},
       eprint = {0704.1220},
 primaryClass = {astro-ph},
       adsurl = {https://ui.adsabs.harvard.edu/abs/2007A&A...468..617S}
}

@ARTICLE{sanna2010a,
       author = {{Sanna}, A. and {Moscadelli}, L. and {Cesaroni}, R. and {Tarchi}, A. and {Furuya}, R.~S. and {Goddi}, C.},
        title = "{VLBI study of maser kinematics in high-mass star-forming regions. I. G16.59-0.05}",
      journal = {\aap},
         year = 2010,
        month = jul,
       volume = {517},
          eid = {A71},
        pages = {A71},
          doi = {10.1051/0004-6361/201014233},
archivePrefix = {arXiv},
       eprint = {1004.2479},
 primaryClass = {astro-ph.GA},
       adsurl = {https://ui.adsabs.harvard.edu/abs/2010A&A...517A..71S}
}

@ARTICLE{sanna2010b,
       author = {{Sanna}, A. and {Moscadelli}, L. and {Cesaroni}, R. and {Tarchi}, A. and {Furuya}, R.~S. and {Goddi}, C.},
        title = "{VLBI study of maser kinematics in high-mass star-forming regions. II. G23.01-0.41}",
      journal = {\aap},
         year = 2010,
        month = jul,
       volume = {517},
          eid = {A78},
        pages = {A78},
          doi = {10.1051/0004-6361/201014234},
archivePrefix = {arXiv},
       eprint = {1004.5578},
 primaryClass = {astro-ph.GA},
       adsurl = {https://ui.adsabs.harvard.edu/abs/2010A&A...517A..78S}
}

@ARTICLE{szymczak2018,
       author = {{Szymczak}, M. and {Olech}, M. and {Sarniak}, R. and {Wolak}, P. and {Bartkiewicz}, A.},
        title = "{Monitoring observations of 6.7 GHz methanol masers}",
      journal = {\mnras},
         year = 2018,
        month = feb,
       volume = {474},
       number = {1},
        pages = {219-253},
          doi = {10.1093/mnras/stx2693},
archivePrefix = {arXiv},
       eprint = {1710.04595},
 primaryClass = {astro-ph.GA},
       adsurl = {https://ui.adsabs.harvard.edu/abs/2018MNRAS.474..219S}
}

@ARTICLE{rosero2016,
       author = {{Rosero}, V. and {Hofner}, P. and {Claussen}, M. and {Kurtz}, S. and {Cesaroni}, R. and {Araya}, E.~D. and {Carrasco-Gonz{\'a}lez}, C. and {Rodr{\'\i}guez}, L.~F. and {Menten}, K.~M. and {Wyrowski}, F. and {Loinard}, L. and {Ellingsen}, S.~P.},
        title = "{Weak and Compact Radio Emission in Early High-mass Star-forming Regions. I. VLA Observations}",
      journal = {\apjs},
         year = 2016,
        month = dec,
       volume = {227},
       number = {2},
          eid = {25},
        pages = {25},
          doi = {10.3847/1538-4365/227/2/25},
archivePrefix = {arXiv},
       eprint = {1609.03269},
 primaryClass = {astro-ph.GA},
       adsurl = {https://ui.adsabs.harvard.edu/abs/2016ApJS..227...25R}
}

@INPROCEEDINGS{McMullin07,
       author = {{McMullin}, J.~P. and {Waters}, B. and {Schiebel}, D. and {Young}, W. and {Golap}, K.},
        title = "{CASA Architecture and Applications}",
    booktitle = {Astronomical Data Analysis Software and Systems XVI},
         year = 2007,
       editor = {{Shaw}, R.~A. and {Hill}, F. and {Bell}, D.~J.},
       series = {Astronomical Society of the Pacific Conference Series},
       volume = {376},
        month = oct,
        pages = {127},
       adsurl = {https://ui.adsabs.harvard.edu/abs/2007ASPC..376..127M}
}

@ARTICLE{tanabe2023,
       author = {{Tanabe}, Yoshihiro and {Yonekura}, Yoshinori},
        title = "{A flare of 6.668 GHz methanol maser in G23.389+0.185}",
      journal = {The Astronomer's Telegram},
         year = 2023,
        month = oct,
       volume = {16303},
        pages = {1},
       adsurl = {https://ui.adsabs.harvard.edu/abs/2023ATel16303....1T}
}

@ARTICLE{devilliers2015,
       author = {{de Villiers}, H.~M. and {Chrysostomou}, A. and {Thompson}, M.~A. and {Urquhart}, J.~S. and {Breen}, S.~L. and {Burton}, M.~G. and {Ellingsen}, S.~P. and {Fuller}, G.~A. and {Pestalozzi}, M. and {Voronkov}, M.~A. and {Ward-Thompson}, D.},
        title = "{6.7-GHz methanol maser associated outflows: an evolutionary sequence}",
      journal = {\mnras},
         year = 2015,
        month = may,
       volume = {449},
       number = {1},
        pages = {119-128},
          doi = {10.1093/mnras/stv173},
archivePrefix = {arXiv},
       eprint = {1501.06589},
 primaryClass = {astro-ph.SR},
       adsurl = {https://ui.adsabs.harvard.edu/abs/2015MNRAS.449..119D}
}

@ARTICLE{1980AA....89..377C,
       author = {{Clark}, B.~G.},
        title = "{An efficient implementation of the algorithm 'CLEAN'}",
      journal = {\aap},
         year = 1980,
        month = sep,
       volume = {89},
       number = {3},
        pages = {377},
       adsurl = {https://ui.adsabs.harvard.edu/abs/1980A&A....89..377C}
}

@INPROCEEDINGS{wolak2024,
       author = {{Wolak}, P. and {Szymczak}, M. and {Bartkiewicz}, A. and {Durjasz}, M. and {Kobak}, A. and {Olech}, M.},
        title = "{Torun methanol maser monitoring program}",
    booktitle = {Cosmic Masers: Proper Motion Toward the Next-Generation Large Projects},
         year = 2024,
       editor = {{Hirota}, Tomoya and {Imai}, Hiroshi and {Menten}, Karl and {Pihlstr{\"o}m}, Ylva},
       series = {IAU Symposium},
       volume = {380},
        month = jan,
        pages = {264-265},
          doi = {10.1017/S1743921323002946},
       adsurl = {https://ui.adsabs.harvard.edu/abs/2024IAUS..380..264W}
}

@ARTICLE{molinari2016,
       author = {{Molinari}, S. and {Schisano}, E. and {Elia}, D. and {Pestalozzi}, M. and {Traficante}, A. and {Pezzuto}, S. and {Swinyard}, B.~M. and {Noriega-Crespo}, A. and {Bally}, J. and {Moore}, T.~J.~T. and {Plume}, R. and {Zavagno}, A. and {di Giorgio}, A.~M. and {Liu}, S.~J. and {Pilbratt}, G.~L. and {Mottram}, J.~C. and {Russeil}, D. and {Piazzo}, L. and {Veneziani}, M. and {Benedettini}, M. and {Calzoletti}, L. and {Faustini}, F. and {Natoli}, P. and {Piacentini}, F. and {Merello}, M. and {Palmese}, A. and {Del Grande}, R. and {Polychroni}, D. and {Rygl}, K.~L.~J. and {Polenta}, G. and {Barlow}, M.~J. and {Bernard}, J. -P. and {Martin}, P.~G. and {Testi}, L. and {Ali}, B. and {Andr{\'e}}, P. and {Beltr{\'a}n}, M.~T. and {Billot}, N. and {Carey}, S. and {Cesaroni}, R. and {Compi{\`e}gne}, M. and {Eden}, D. and {Fukui}, Y. and {Garcia-Lario}, P. and {Hoare}, M.~G. and {Huang}, M. and {Joncas}, G. and {Lim}, T.~L. and {Lord}, S.~D. and {Martinavarro-Armengol}, S. and {Motte}, F. and {Paladini}, R. and {Paradis}, D. and {Peretto}, N. and {Robitaille}, T. and {Schilke}, P. and {Schneider}, N. and {Schulz}, B. and {Sibthorpe}, B. and {Strafella}, F. and {Thompson}, M.~A. and {Umana}, G. and {Ward-Thompson}, D. and {Wyrowski}, F.},
        title = "{Hi-GAL, the Herschel infrared Galactic Plane Survey: photometric maps and compact source catalogues. First data release for the inner Milky Way: +68{\textdegree} {\ensuremath{\geq}} l {\ensuremath{\geq}} -70{\textdegree}}",
      journal = {\aap},
         year = 2016,
        month = jul,
       volume = {591},
          eid = {A149},
        pages = {A149},
          doi = {10.1051/0004-6361/201526380},
archivePrefix = {arXiv},
       eprint = {1604.05911},
 primaryClass = {astro-ph.GA},
       adsurl = {https://ui.adsabs.harvard.edu/abs/2016A&A...591A.149M}
}

@ARTICLE{Wright2010,
       author = {{Wright}, Edward L. and {Eisenhardt}, Peter R.~M. and {Mainzer}, Amy K. and {Ressler}, Michael E. and {Cutri}, Roc M. and {Jarrett}, Thomas and {Kirkpatrick}, J. Davy and {Padgett}, Deborah and {McMillan}, Robert S. and {Skrutskie}, Michael and {Stanford}, S.~A. and {Cohen}, Martin and {Walker}, Russell G. and {Mather}, John C. and {Leisawitz}, David and {Gautier}, Thomas N., III and {McLean}, Ian and {Benford}, Dominic and {Lonsdale}, Carol J. and {Blain}, Andrew and {Mendez}, Bryan and {Irace}, William R. and {Duval}, Valerie and {Liu}, Fengchuan and {Royer}, Don and {Heinrichsen}, Ingolf and {Howard}, Joan and {Shannon}, Mark and {Kendall}, Martha and {Walsh}, Amy L. and {Larsen}, Mark and {Cardon}, Joel G. and {Schick}, Scott and {Schwalm}, Mark and {Abid}, Mohamed and {Fabinsky}, Beth and {Naes}, Larry and {Tsai}, Chao-Wei},
        title = "{The Wide-field Infrared Survey Explorer (WISE): Mission Description and Initial On-orbit Performance}",
      journal = {\aj},
         year = 2010,
        month = dec,
       volume = {140},
       number = {6},
        pages = {1868-1881},
          doi = {10.1088/0004-6256/140/6/1868},
archivePrefix = {arXiv},
       eprint = {1008.0031},
 primaryClass = {astro-ph.IM},
       adsurl = {https://ui.adsabs.harvard.edu/abs/2010AJ....140.1868W}
}
\bibliographystyle{aasjournalv7}

\end{document}